\documentclass{aa}
\usepackage[varg]{txfonts}

\usepackage{graphicx}
\usepackage{tabularx}

\usepackage{natbib,twoopt}
\usepackage[breaklinks=true]{hyperref} 
\hypersetup{
  colorlinks=true,   
  urlcolor=blue,     
  linkcolor=blue     
}

\bibpunct{(}{)}{;}{a}{}{,}             
\makeatletter
  \newcommandtwoopt{\citeads}[3][][]{\href{http://adsabs.harvard.edu/abs/#3}%
    {\def\hyper@linkstart##1##2{}%
     \let\hyper@linkend\@empty\citealp[#1][#2]{#3}}}
  \newcommandtwoopt{\citepads}[3][][]{\href{http://adsabs.harvard.edu/abs/#3}%
    {\def\hyper@linkstart##1##2{}%
     \let\hyper@linkend\@empty\citep[#1][#2]{#3}}}
  \newcommandtwoopt{\citetads}[3][][]{\href{http://adsabs.harvard.edu/abs/#3}%
    {\def\hyper@linkstart##1##2{}%
     \let\hyper@linkend\@empty\citet[#1][#2]{#3}}}
  \newcommandtwoopt{\citeyearads}[3][][]%
    {\href{http://adsabs.harvard.edu/abs/#3}
    {\def\hyper@linkstart##1##2{}%
     \let\hyper@linkend\@empty\citeyear[#1][#2]{#3}}}
 \renewcommand*\aa@pageof{, page \thepage{} of \pageref*{LastPage}} 
\makeatother

\usepackage{booktabs}
\usepackage{epstopdf}

\usepackage[capitalise]{cleveref}

\usepackage{bm}
\usepackage{physics}

\begin{document}

\title{Effect of optically thin cooling curves on condensation formation: Case study using thermal instability}

\titlerunning{Effect of cooling curves on condensation formation}

\author{J. Hermans\inst{1}\and R. Keppens\inst{1}}

\institute{$^1$ Centre for mathematical Plasma-Astrophysics, Celestijnenlaan 200B, 3001 Leuven, KU Leuven, Belgium} 

\date{Received date /
Accepted date }
 
\abstract{Non-gravitationally induced condensations are observed in many astrophysical environments. In solar physics, common phenomena are coronal rain and prominences. These structures are formed due to energy loss by optically thin radiative emission. Instead of solving the full radiative transfer equations, precomputed cooling curves are typically used in numerical simulations. In the literature, a wide variety of cooling curves exist, and they are quite often used as unquestionable ingredients.}{We here determine the effect of the optically thin cooling curves on the formation and evolution of condensations. We also investigate the effect of numerical settings. This includes the resolution and the low-temperature treatment of the cooling curves, for which the optically thin approximation is not valid.}{We performed a case study using thermal instability as a mechanism to form in situ condensations. We compared 2D numerical simulations with different cooling curves using interacting slow magnetohydrodynamic (MHD) waves as trigger for the thermal instability. Furthermore, we discuss a bootstrap measure to investigate the far non-linear regime of thermal instability. In the appendix, we include the details of all cooling curves implemented in MPI-AMRVAC and briefly discuss a hydrodynamic variant of the slow MHD waves setup for thermal instability.}{For all tested cooling curves, condensations are formed. The differences due to the change in cooling curve are twofold. First, the growth rate of the thermal instability is different, leading to condensations that form at different times. Second, the morphology of the formed condensation varies widely. After the condensation forms, we find fragmentation that is affected by the low-temperature treatment of the cooling curves. Condensations formed using cooling curves that vanish for temperatures lower than 20\,000 K appear to be more stable against dynamical instabilities. We also show the need for high-resolution simulations. The bootstrap procedure allows us to continue the simulation into the far non-linear regime, where the condensation fragments dynamically align with the background magnetic field. The non-linear regime and fragmentation in the hydrodynamic case differ greatly from the low-beta MHD case.}{We advocate the use of modern cooling curves, based on accurate computations and current atomic parameters and solar abundances. Our bootstrap procedure can be used in future multi-dimensional simulations to study fine-structure dynamics in solar prominences.}


\maketitle

\section{Introduction}\label{Sec: Intro}

Optically thin radiative emission is an important physical process. The energy loss due to recombination of ions can cause a significant decrease in the temperature of a plasma. Hence it plays a key role in the formation of condensations in various fields of astrophysics.

Examples of condensed structures in solar physics are prominences and coronal rain. Prominences are cool and dense structures that are suspended in the hot and tenuous corona by the magnetic field \citepads{2014LRSP...11....1P}. Coronal rain is a more transient and recurrent phenomenon, with a lifetime of only minutes to days. After being formed at the top of magnetic loops, the dense blobs rain down along field lines in cycles \citepads{2020PPCF...62a4016A, 2020A&A...633A..11F}. In the interstellar medium, turbulent flows can lead to the formation of dense molecular clouds, from which stars can form in turn \citepads{2005A&A...433....1A}. Condensations are also observed  as neutral-atomic and molecular outflows in galaxies \citepads{2020A&ARv..28....2V} and as filaments in the intergalactic medium \citepads{2018arXiv181112147S}.

The modelling of the thermodynamics of such phenomena is not straightforward. Solving the full radiative transfer equations alongside the magnetohydrodynamics (MHD) equations is possible. An example is the Bifrost code \citepads{2011A&A...531A.154G}, which has recently been used to study coronal rain formation \citepads{2020A&A...639A..20K}. However, this is computationally demanding for multidimensional simulations.

Under several assumptions, especially when dealing only with optically thin radiation, the density and temperature dependencies of the radiative energy loss can be separated \citepads{2003adu..book.....D}. The temperature dependence is typically calculated in advance and tabulated as a cooling curve. A large variety of cooling curves exists in the literature \citepads[see e.g.]{1969ApJ...157.1157C,1974SoPh...35..123H,1978ApJ...220..643R,2008ApJ...689..585C,2009A&A...498..915D,2009A&A...508..751S}, and in our appendix, we provide an overview of the most frequently employed cooling curves, which we have all implemented in our open-source software MPI-AMRVAC \citepads{2012JCoPh.231..718K,2014ApJS..214....4P,2018ApJS..234...30X,2020arXiv200403275K}.  Differences arise due to different assumptions in the calculation, varying atomic physics parameter, and the solar abundances that are used \citepads{1978ApJ...220..643R}. The effect of improved atomic physics has been proven, for example, in the field of asteroseismology. Advances in opacity tables for elements such as iron and nickel have led to an explanation of the excitation of pulsations in B-type stars \citepads{1994IAUS..162...55D}.

In this work we study the effect of varying the optically thin cooling curve on the formation of condensations. Various physical processes can lead to condensations. We form in situ condensations by thermal instability (TI), first described by \citetads{1953ApJ...117..431P} and \citetads{1965ApJ...142..531F}. The theory was recently revisited by \citetads{2019A&A...624A..96C} and was simulated in an idealised setup. This setup of a local typical solar coronal volume showed that condensations arise naturally due to optically thin cooling and a perturbation, independent of the origin of the perturbation. \citetads{2020A&A...636A.112C} included the effects of anisotropic thermal conduction and showed that the fragmentation of the condensations and the magnetic field lines is misaligned. This complicates the interpretation of observed fine structure in solar prominences, for instance. \citetads{2012A&A...540A...7S} investigated the stability of thermal modes in cool prominence plasmas for three typical cooling curves.

Thermal instability is a very likely mechanism to form solar prominences, as was recently shown in numerical simulations by \citetads{2021A&A...646A.134J}. A review by \citetads{2010SSRv..151..333M} discussed a variety of mechanisms to form prominences, not necessarily invoking thermal instability. The formation of other condensed structures, such as filaments in the interstellar and intergalatic medium are also being explained by thermal instability \citepads[see e.g. ]{1972ApJ...178..143C,2010ApJ...720..652S}. Recently, \citetads{2020MNRAS.492.4484F} revisited thermal instability in the context of shock-induced condensation formation in the interstellar medium as a process to generate density inhomogeneities without relying on self-gravity. \citetads{2021MNRAS.505.5238J} studied cloud formation in the interstellar medium by thermal instability using Gaussian random fields as initial conditions and including magnetic fields, thermal conduction, and viscosity. They also discussed fragmentation processes for hydrodynamic clouds. Furthermore, \citetads{2019ApJ...875..158W} investigated the possible regimes of thermal instability in hydrodynamics. They discussed the stability of the entropy and acoustic modes between the classical isobaric and isochoric limits. They also performed 1D numerical simulations of the non-linear evolution of the thermal instability and discussed fragmentation mechanisms. With respect to this, it has been shown that clumps in active galactic nucleus (AGN) winds can be explained by thermal instability \citepads{2020ApJ...893L..34D,2021ApJ...914...62W}.  It should be noted that the effect of typical AGN net cooling curves, which depend on the temperature and photoionisation parameters, on the stability of plasma has been studied in the recent past \citepads[see e.g.][]{2008MNRAS.384L..24C,2009MNRAS.393...83C,2012MNRAS.422..637C}.

The results of this case study using thermal instability might also be applicable to other mechanisms that rely heavily on optically thin radiative cooling. In thermal non-equilibrium models (TNE), evaporation and condensation cycles can create coronal rain \citepads{2020PPCF...62a4016A}. A difference between TNE and thermal instability is that most studies on thermal instability start from a thermal equilibrium, which is then perturbed \citepads{2019SoPh..294..173K}. In TNE studies, the focus lies instead on time-varying conditions that may never be in perfect thermal equilibrium, and the non-linear evolution shows runaway cooling, denoting the fast decrease in temperature due to optically thin cooling. In both processes the cooling, and thus the formation of condensations, is dependent on the cooling curves.

In this work we use a simplified setup of thermal instability with solar coronal conditions, following the method of \citetads{2019A&A...624A..96C} and \citetads{2020A&A...636A.112C}. The initial thermal equilibrium will be perturbed by slow MHD waves. Slow MHD waves are commonly observed in the solar corona \citepads[see][and references therein]{2020arXiv201208802B} and are therefore a good choice for the perturbation.

In \cref{Sec: coolingcurves} we discuss the optically thin cooling curves. In \cref{Sec: Setup} we present the setup, initial conditions, and numerical methods. The benchmark simulation is discussed in \cref{Sec: benchmark}, and a comparison is made to the setup and results of \citetads{2020A&A...636A.112C}. The effect of several numerical settings, such as the resolution and the detailed low-temperature treatment, are investigated in \cref{Sec: effectNum}. We compare simulations with different cooling curves in \cref{Sec: Dif cc}. Using the insights gained in preceding sections, we follow the evolution of the thermal instability far into the non-linear regime in \cref{Sec: Extending}. We present a summary of the results and discussion in \cref{Sec: sum-dis}. \cref{App: ccs} gives references to all cooling curves currently available in the MPI-AMRVAC software we use. \cref{App: HD} contrasts the low plasma beta MHD case studied in this manuscript with a related hydrodynamics case.

\section{Optically thin radiative cooling curves}\label{Sec: coolingcurves}

The energy loss due to optically thin radiative cooling is typically written as
\begin{equation}\label{Eq: dedtunit}
    \pdv{e}{t} = -n_i n_e\Lambda(T),
\end{equation}
where $n_e$, $n_i$, and $T$ are the local electron and ion number density and the temperature of a plasma, respectively. The variable $e$ denotes the internal or total energy density, and the function $\Lambda(T)$ is the predefined cooling curve that is also called the cooling rate. This equation is also commonly written in normalised variables as
\begin{equation}\label{Eq: dedt}
    \pdv{\bar{e}}{\bar{t}} = -\bar{\rho}^2\bar{\Lambda}(\bar{T}),
\end{equation}
where $\bar{\rho}$ is the normalised local density. All variables in this equation are normalised, in contrast to the variables in \cref{Eq: dedtunit}, as denoted by the overlying bar. To normalise the equations, we use the unit values of the quantities of the initial state given in \cref{Tab: initial_state}. All other unit quantities can be derived from the chosen three quantities and are also given in \cref{Tab: initial_state}. As an example, the normalisation of the density from $\rho$ to $\bar{\rho}$ and the definition of $\rho_u$ is given by
\begin{equation}
     \bar{\rho} = \frac{\rho}{\rho_{u}} =  \frac{\rho}{(1+4A_{He})m_{p}n_u},
\end{equation}
where $\rho_{u}$, $n_u$, $m_p$, and $A_{He}$ are the unit density, the unit number density from \cref{Tab: initial_state}, the proton mass, and the assumed hydrogen-helium abundance ratio, respectively. We used a hydrogen-helium abundance ratio of 0.1. The factor four arises from mass conservation, that is, the fact that a helium ion has four times more mass than a hydrogen ion. The cooling curve can be normalised as
\begin{equation}
    \bar{\Lambda} = \frac{\Lambda}{\Lambda_u} = \Lambda\frac{t_u n_{u}^2 (1+2A_{He})}{p_u},
\end{equation}
where $\Lambda_u$, $t_u$, and $p_u$ are the unit cooling rate, unit time, and unit pressure, respectively. The last two can be derived similarly. The factor two arises from charge conservation. Unless stated otherwise, all equations are written in normalised variables, hence we drop the overlying bar. 

The energy loss is due to emission, such as line and continuum radiation and bremsstrahlung. It is typically calculated under the assumption of collisional ionisation equilibrium (CIE), meaning that collisional ionisation balances the recombination processes. A necessary condition is that the plasma is optically thin, so that all radiation produced within is able to escape and/or is fully ionised so that photoionisation can be ignored. Furthermore, it is commonly assumed that all excited ions return to their ground state faster than the time between collisions is long, so that only the ground state of each ionic species is populated \citepads{2003adu..book.....D}. However, this is not a necessary assumption. Cooling curves can also be constructed by solving the full collisional-radiative rate equation, that is, calculating the populations of all the excited levels \citepads[see e.g.][]{2008ApJ...689..585C}. Under these assumptions, the density and temperature dependences can be separated. A cooling curve thus represents the dependence of the energy loss on the local temperature of the medium.

In addition to the specific method adopted within the computation, the accuracy of a cooling curve depends on two other things: the accuracy of the atomic physics parameters, and the accuracy of the assumed solar abundances \citepads{1978ApJ...220..643R}. The effect of several parameters, such as solar abundances, ion fractions, electron densities, and transition probabilities, has been investigated \citepads{1999A&A...347..401L}. The abundances seem to be particularly important, although other parameters involved can also have a strong effect on a cooling curve.

We compared five different radiative cooling curves for optically thin cooling that were used in past or present solar physics research. We briefly describe these cooling curves in this section. In some works that studied solar phenomena, such as flares and prominences \citepads[see e.g.][]{1991SoPh..135..361F,2015ApJ...806..115K} or thermal instability in molecular cloud formation \citepads[e.g.][]{2010ApJ...720..652S}, cooling curves were modified for numerical reasons, such as to speed up the growth of the thermal instability or to increase the characteristic length scale of thermal conduction. We did not investigate these particular changes, but rather compare the effects of changing the entire cooling curves.
 
Before we discuss the five cooling curves, we address the numerical implementation of a cooling curve.
The $\Lambda(T)$ prescriptions typically come in two forms: interpolatable tables, or piece-wise power laws. As the name suggests, an interpolatable table is a set of data points, typically 50 to 100 points, that can be interpolated with a higher temperature-resolution to a more extended table at the start of a simulation. During runtime, the values of the cooling rate at a given temperature are selected from this table. The piece-wise power laws are analytical expressions. Their cooling rate is just calculated based on the given temperature. They are coarser than the interpolatable tables because they typically consist of fewer than ten segments. The piece-wise power laws have the advantage that they can easily be used in analytic calculations, such as in 0D models of coronal rain \citepads{1995ApJ...439.1034C}. They have also been used in early studies of magneto-thermal instabilities in non-uniform coronal plasmas \citepads{1992SoPh..140..317V}, which have recently been revisited with the linear MHD code Legolas \citepads{2020ApJS..251...25C}.

We now describe the five cooling curves. For details about their calculations, we refer to the accompanying papers. \cref{App: ccs} gives an overview of the various available curves, with the original references. The first three cooling curves are interpolatable tables.
The first cooling curve is the SPEX\_DM table as described in \citetads{2009A&A...508..751S}. The main temperature regime, 10$^{3.8}$ to 10$^{8.16}$ K, was calculated with the SPEX package \citepads{2000adnx.conf..161K}. The solar abundances of \citetads{1989GeCoA..53..197A} were used. For temperatures below 10\,000 K, the radiative cooling is approximated by the cooling curve first described by \citetads{1972ARA&A..10..375D}, but implemented as presented by \citetads{2009A&A...508..751S}. The cooling of this \_DM part is due to the collisional excitation of singly charged ions of O, C, N, Si, Fe, Ne, and S with thermal electrons and neutral hydrogen. There is also a contribution by collisions between neutral hydrogen with electrons. An ionisation fraction of 10$^{-3}$ is assumed for the low-temperature cooling curve. The SPEX\_DM table is also the reference cooling curve in this work.

The Colgan\_DM cooling curve is given in \citetads{2008ApJ...689..585C} as their `DW ground' model. It is constructed assuming that only the ground state of ions is populated. The atomic and collisional calculations were performed with the Los Alamos plasma kinetics code ATOMIC. The abundances of \citetads{1998SSRv...85..161G}, modified by \citetads{2003SSRv..107..665F}, were used. This table is extended for low temperatures by the \_DM part described above.

The third cooling curve, Dere\_corona\_DM, is described and given in \citetads{2009A&A...498..915D}. This table was constructed with version 6 of the CHIANTI atomic database \citepads{1997A&AS..125..149D}. They used the abundances from \citetads{1992PhyS...46..202F}. This table is also extended by the \_DM part for low temperatures.

The first piece-wise power law is the cooling curve by \citetads{1978ApJ...220..643R}. It is actually an analytical fit to an unpublished interpolatable table by J. Raymond. It uses the solar coronal abundances that were accepted at that time, but the specific details are not known. We used a version that is extended for high and low temperatures by \citetads{1982soma.book.....P}.

The last cooling curve is the piece-wise power law by \citetads{2008ApJ...682.1351K}. Although the cooling curve was described in 2008, it was based on much older physical data and calculations. The atomic physics calculations were provided by J. Raymond by means of private communication in 1994. The coronal abundances of \citetads{1985ApJS...57..173M} were modified by a factor of two.

\begin{figure}[htbp]
  \resizebox{\hsize}{!}{\includegraphics{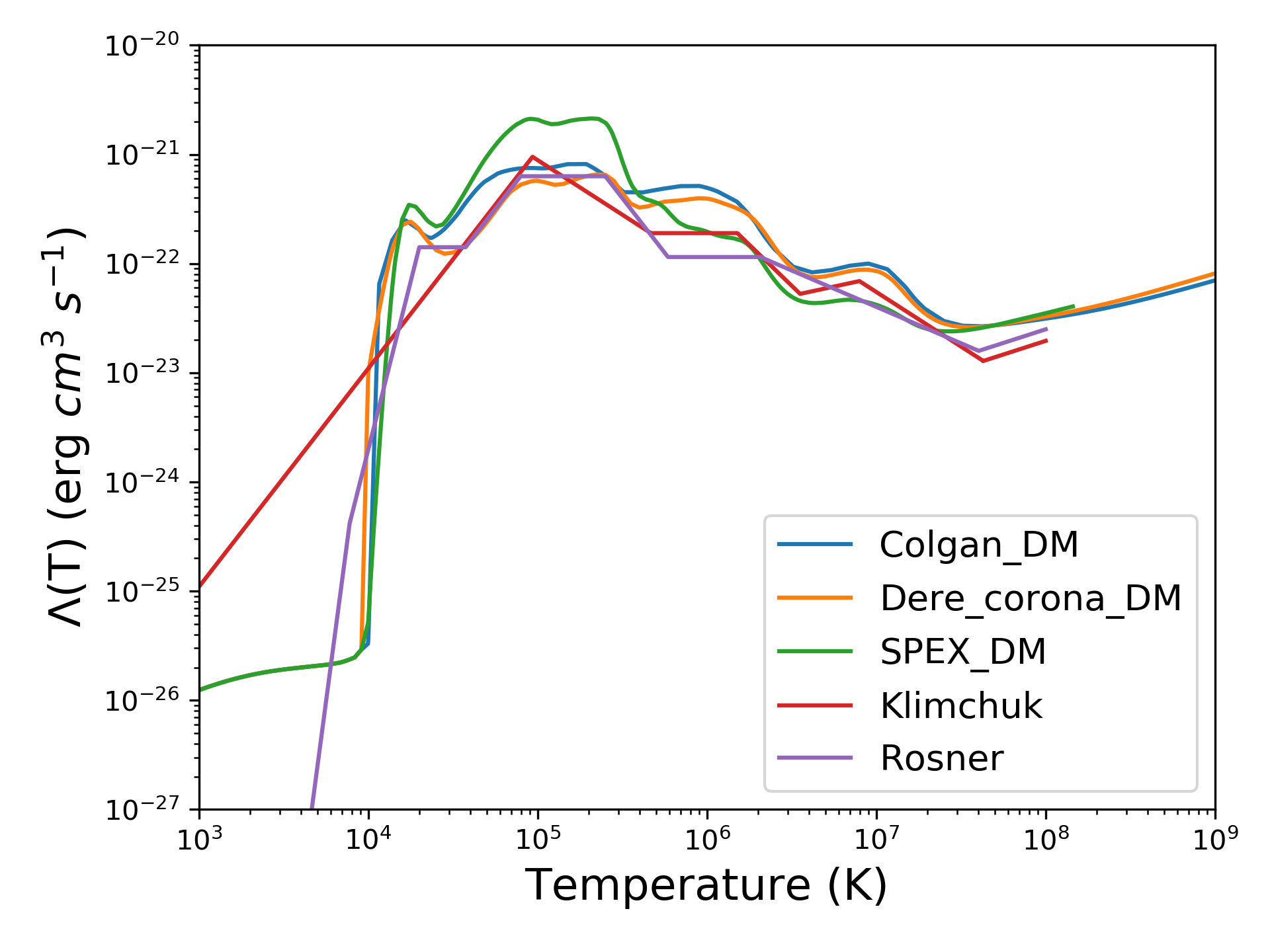}}
  \caption{Five cooling curves that we compare.}
  \label{Fig: Coolingcurves}
\end{figure}

These five cooling curves are depicted in \cref{Fig: Coolingcurves} on a log-log scale. There are many differences between the curves. The two tables that differ the least are the Colgan\_DM and the Dere\_corona\_DM. This is expected because they both use current values for the abundances. As discussed by \citetads{2009A&A...498..915D}, this might be due to the different amount of atomic fine structure that is used. The piece-wise power laws are much coarser and hence do not represent all variations in this detail. A striking difference between the Colgan\_DM and Dere\_corona\_DM and the other three curves is that the last three have a much lower cooling rate from 10$^6$ K and up. This is mostly because these three cooling curves used older solar abundances. The difference is almost the same as between solar corona and photospheric abundances, as reported by \citetads{2009A&A...498..915D}. In the 1980s, it became known that the quiet-Sun upper atmosphere can have a fourfold increase in abundance of low first-ionisation potential elements, such as Fe, Si, and Mg, with respect to the solar photosphere \citepads{1992PhyS...46..202F}. It might therefore be argued that cooling curves based on older abundances are less representative of the optically thin radiative energy loss in the solar corona. Nevertheless, because all five curves have been (and still are) used in contemporary MHD simulations of the solar corona, we here set forth to study a simple idealised setup that truly isolates their effect on the formation of condensations.

Several comparisons can be made without multidimensional MHD simulations. One of them is the linear cooling timescale $\tau_r$, also called radiative timescale. It describes the time it takes for a parcel of plasma of a given density and temperature to cool. For a given temperature $T,$ it can be written as
\begin{equation}\label{Eq: tau_r}
    \tau_r(T) = \frac{T}{(\gamma-1)\rho \Lambda(T)},
\end{equation}
\noindent where $\rho$ and $\Lambda$ are the density and cooling rate, respectively. We take $\gamma$, the ratio of specific heats, to be 5/3 by assuming a monatomic ideal gas.

\begin{figure}[htbp]
  \resizebox{\hsize}{!}{\includegraphics{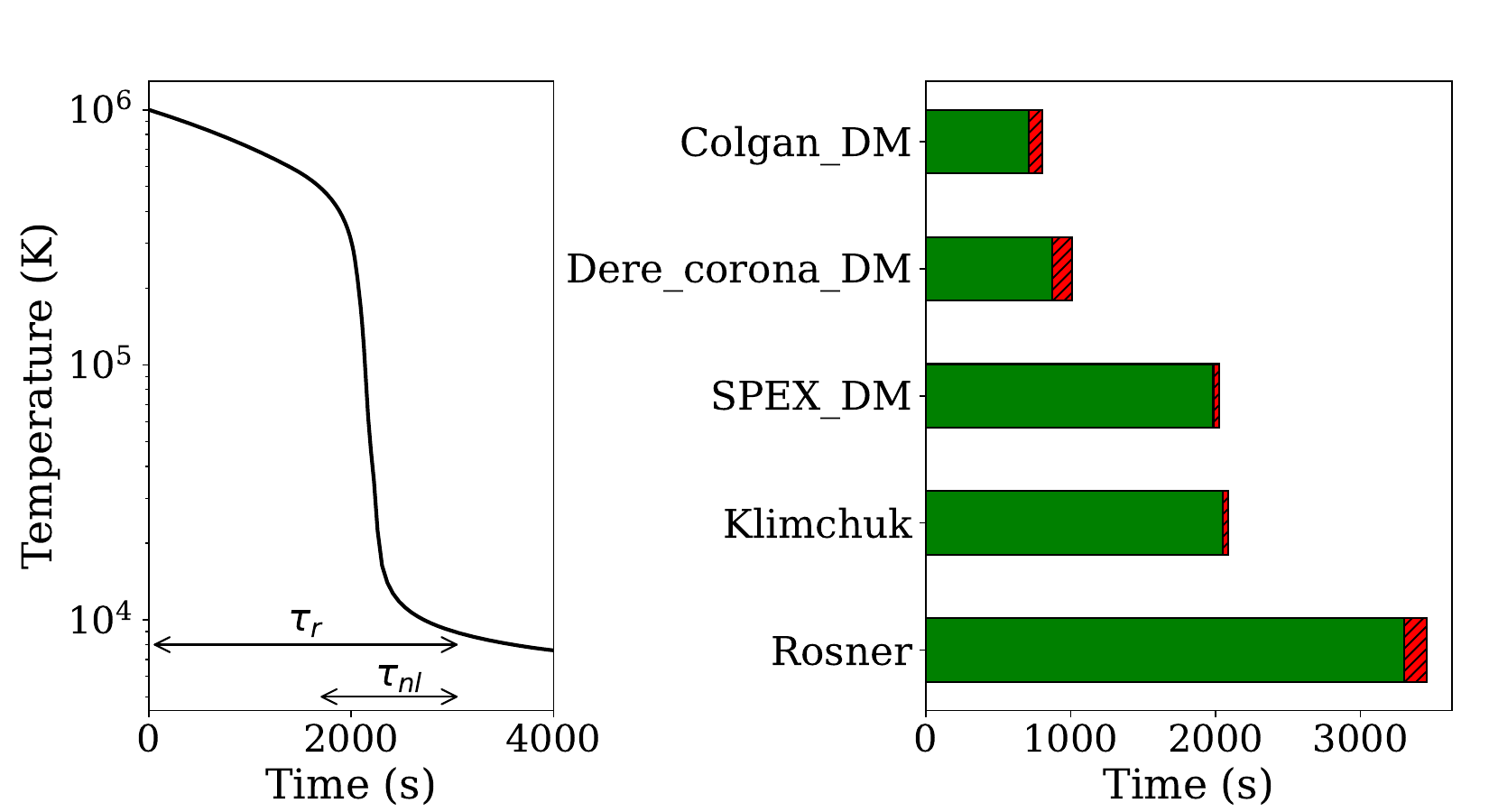}}
  \caption{\textit{Left}: General thermal evolution of thermal instability and its timescales. \textit{Right}: Linear and non-linear timescales of five cooling curves. The linear timescale is depicted by the length of the full bar, and the non-linear part is represented by the red part.}
  \label{Fig: Timescales}
\end{figure}

We can also define a nonlinear radiative timescale as
\begin{equation}\label{Eq: tau_nl}
    \tau_{nl}(T) = \frac{T_{max} \Lambda(T)}{T\Lambda(t_{max})}\tau_r(T),
\end{equation}
\noindent where $T_{max}$ is the temperature at which the cooling curve has a maximum \citepads{1991SoPh..135..361F}. The physical interpretation of the timescales are depicted in \cref{Fig: Timescales}, which shows the solution of \cref{Eq: dedt} as a function of time for a fixed density and initial temperature. The radiative timescale is the time it takes to cool from a starting temperature, while the non-linear timescale is the time of the non-linear part of the cooling.

For typical solar coronal conditions, that is, a temperature of 10$^6$ K and a density of 2.34~10$^{-15}$ g cm$^{-3}$, these two timescales are shown in \cref{Fig: Timescales} for the five cooling curves. The full length of the bar denotes the radiative timescale, and the red part denotes the non-linear timescale. A first observation is that the length of the timescale corresponds to the cooling rate, as evidently follows from \cref{Eq: tau_r}. There are large differences between the radiative timescales of the different cooling curves. The timescales for the Colgan\_DM and Dere\_corona\_DM are quite similar, which is also seen in the similarity of their cooling curves. This suggests that they will give similar results in numerical simulations. However, in full MHD simulations, other non-linear and dynamical effects will be present. Therefore these values can only be used as a guideline. For example, from this we can assume that numerical simulations using the Rosner table will take much longer than for the other tables if we are interested in obtaining thermally driven condensations.

\section{Numerical setup}\label{Sec: Setup}

We study the effect of using different cooling curves on condensation formation by thermal instability. To initiate the thermal instability, we follow the method from \citetads{2019A&A...624A..96C} and \citetads{2020A&A...636A.112C}. 

\subsection{Governing equations}\label{Subsec: goveq}

We make use of the MPI-parallelised Adaptive Mesh Refinement Versatile Advection Code (MPI-AMRVAC)\footnote{Open source at \url{http://amrvac.org}} \citepads{2012JCoPh.231..718K,2014ApJS..214....4P,2018ApJS..234...30X,2020arXiv200403275K} to numerically solve the following set of normalised MHD equations:
\begin{align}\label{Eq: drhodt}
    &\pdv{\rho}{t} + \nabla \cdot (\bm{v}\rho) =0,\\
    &\pdv{\rho\bm{v}}{t} +\nabla \cdot ( \rho\bm{v}\bm{v} + p_{tot}\bm{I} - \bm{B}\bm{B}) = 0,\\
    &\pdv{e}{t} + \nabla \cdot (e\bm{v} + p_{tot}\bm{v} - \bm{B}\bm{B}\cdot\bm{v}) =  - \rho^2 \Lambda(T) + H,\\ 
    &\pdv{\bm{B}}{t} + \nabla \cdot (\bm{vB} - \bm{Bv})=0, \quad \nabla \cdot \bm{B} = 0, \label{Eq: dBdt}
\end{align}

\noindent where $\bm{I}$ is the unit tensor, with the magnetic field in units, where $\mu_0 = 1$. The quantities $\rho$, $T$, $\bm{v}$, and $\bm{B}$ are the density, temperature, velocity, and magnetic field, respectively. The total energy density $e$ is given by
\begin{equation}
    e = \frac{p}{\gamma -1} + \frac{\rho v^2}{2} + \frac{B^2}{2},
\end{equation}

\noindent and the total pressure is given by
\begin{equation}
    p_{tot} = p + \frac{B^2}{2}.
\end{equation}

\noindent The relation between the density and temperature is described by the ideal gas law, thus closing our set of equations. We considered a fully ionised plasma of hydrogen and helium with an abundance ratio of ten hydrogen atoms for each helium atom. This ratio was used in the normalising of the equations and variables.

To keep the setup as simple as possible, we neglected gravity, resistivity, conduction, and other non-adiabatic effects, except for optically thin radiative cooling and a constant background heating, denoted by $H$. The optically thin radiative cooling depends on the local density and cooling rate of the prescribed cooling curves. In order to have an initial state in thermal equilibrium, we took the background heating equal to the radiative loss of the initial state,
\begin{equation}\label{Eq: heating}
    H = \rho_0^2\Lambda(T_0),
\end{equation}
\noindent where $\rho_0$ and $T_0$ are the equilibrium density and temperature, respectively. The background heating is therefore constant.
However, it should be noted that this background heating $H$ is not the heating per unit mass, which is part of the net heat-loss function, denoted by $\mathcal{L}$, as typically used in analytical research of thermal instability and as defined by \citetads{1965ApJ...142..531F}. The net heat-loss function is here given by
\begin{equation}\label{Eq: curlyL}
    \mathcal{L} = \rho \Lambda(T) - \frac{\rho_0^2 \Lambda(T_0)}{\rho}.
\end{equation}
Having the initial state in thermal equilibrium also allows direct comparisons between predicted growth rates of the thermal instability and the onset stages of the condensations, as demonstrated previously by \citetads{2019A&A...624A..96C}.

\subsection{Initial conditions}

To simulate the in situ condensation-forming process by thermal instability, we set up a local 2D coronal volume with periodic boundary conditions. We started from a thermal equilibrium state that was nearly uniform, except for a superposed wave solution. The equilibrium number density and temperature were used together with the characteristic length scale $L$ of the simulation box to normalise the equations and variables. Their values are given in \cref{Tab: initial_state}. The equilibrium values for the density and the temperature are thus equal to unity in normalised notation.

\begin{table}[h!]
\caption{Unit values of the quantities used to normalise the MHD equations and of the initial state and derived unit quantities}              
\label{Tab: initial_state}     
\centering                                 
\begin{tabular}{c c}          
\hline\hline      
Quantity & Value \\ 
\hline                                   
 Number density $n_{u}$ & 1.0 $\cdot$ 10$^{9}$ cm$^{-3}$  \\
 Temperature $T_u$ &  1.0 $\cdot$ 10$^6$ K \\ 
 Length scale $L$ & 1.0 $\cdot$ 10$^9$ cm\medskip\\
\hline\hline
Derived quantity & Value\\
\hline
 Density $\rho_u$ &  2.3 $\cdot$ 10$^{-15}$ g cm$^{-3}$  \\
 Pressure $p_u$ &  3.2 $\cdot$ 10$^{-1}$ dyn cm$^{-2}$\\
 Velocity $v_u$ & 1.2 $\cdot$ 10$^{7}$ cm s$^{-1}$\\
 Time $t_u$ & 8.6 $\cdot$ 10 s\\
 Magnetic field $B_u$ & 2.0 G\\
 Cooling rate $\Lambda_u$ & 3.1 $\cdot$ 10$^{-21}$ erg cm$^{3}$ s$^{-1}$\\
 \hline
\end{tabular}
\end{table}

The thermal mode, which is the entropy mode, can be excited directly, but here we followed \citetads{2019A&A...624A..96C} and used two interacting slow MHD modes to perturb the thermal equilibrium and facilitate the thermal instability. The slow waves become damped due to the radiative cooling, while the thermal mode is unstable for the physical conditions in \cref{Tab: initial_state}, as was shown by \citetads{2019A&A...624A..96C}. The isobaric and isochoric instability criteria are both satisfied for the given initial conditions. Because the plasma has been changed by the interacting, damped slow wave propagation, it is not trivial to categorise the linear thermal mode as a pure isobaric or isochoric mode, although the observed growth rate matches analytic predictions reasonably well \citepads{2019A&A...624A..96C}.

We excited one slow wave along each axis. The wave number, $k$, was equal to $2\pi$ divided by the domain length $L$ to satisfy the periodic boundary conditions by having one full oscillation within the domain. 

The direction of the background magnetic field was taken to be along the diagonal of the square domain. This was not the case in \citetads{2020A&A...636A.112C}, who allowed it to vary for different simulations. The implications and differences are discussed in \cref{Sec: benchmark}. The background magnetic field had a strength of 10 G, and the medium had a plasma beta of approximately 0.08, which are typical for prominences in the solar corona \citepads{2018LRSP...15....7G}.

The eigenfrequency $\omega$ of the slow modes does not include the effect of the cooling curve on the initial condition. We used the adiabatic slow MHD eigenfrequency as can be found in any textbook on MHD \citepads[see e.g.][]{GoedbloedJP2019MoLa}. It can be calculated as follows:
\begin{equation}\label{Eq: wr}
    \omega_s = k \sqrt{\frac{1}{2}(v_A^2 + c^2) - \frac{1}{2}\sqrt{(v_A^2 + c^2)^2 -4c^2v_A^2\cos^2{\theta}} },
\end{equation}
\noindent where the sound- and Alfv\'en speeds are defined by  
\begin{equation}
    c^2 = \frac{\gamma p_0}{\rho_0} \quad \text{and} \quad v_A^2 = \frac{B^2_0}{\rho_0}.
\end{equation}
\noindent The angle $\theta$ is defined between $\bm{k}$ and $\bm{B}$ and had a value of 45$\degr$ for all simulations in this work because we set up the magnetic field along the diagonal of the square domain and then adopted one slow mode travelling along the $x$ -axis and one slow mode travelling along the $y$ -axis.

\begin{figure}[htbp]
  \resizebox{\hsize}{!}{\includegraphics{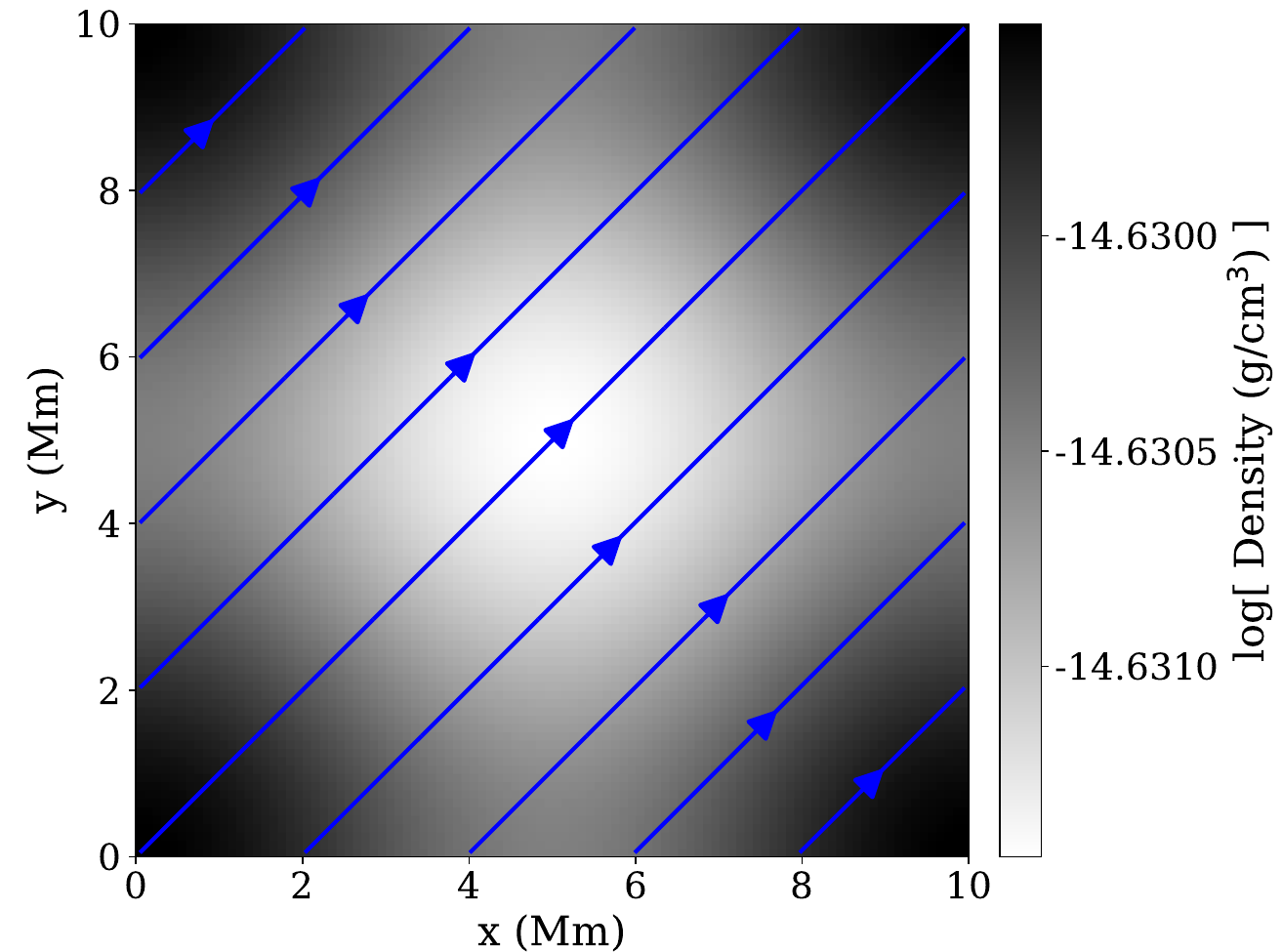}}
  \caption{Density view of the initial condition of two interacting slow MHD waves perturbing the thermal equilibrium. Magnetic field lines are superimposed in blue.}
  \label{Fig: initial config}
\end{figure}

Lastly, the slow MHD waves were initiated by perturbation of the thermal equilibrium. The eigenfunctions were derived and described by \citetads{2020A&A...636A.112C}. The initial velocity perturbation was taken as a plane wave and given by 
\begin{equation}
    v_x = A[\cos{(k_x x)} + i \sin{(k_x x)}],
\end{equation}
\noindent with $A$ an amplitude taken to be 10$^{-3}$. The velocity perturbation along the $y$ -axis is similar. The real part of complex eigenfunctions was used. A density view of the initial state, with the magnetic field lines superimposed, is given in \cref{Fig: initial config}.

\subsection{Numerical methods}

The set of equations, \crefrange{Eq: drhodt}{Eq: dBdt}, was solved on a 2D Cartesian grid using a strong stability preserving five-step fourth-order Runga-Kutta method \citepads{Spiteri2002ANC} and an HLL flux scheme, first described by \citetads{1983HLL}. To compute fluxes at the cell edges, cell centers values were reconstructed where we used a second-order symmetric total variation diminishing (TVD) limiter, denoted `woodward', \citepads{1977JCoPh..23..276V,1984JCoPh..54..115W}. We used a Courant number of 0.8 to limit the time step used by the explicit Runga-Kutta method according to the Courant–Friedrichs–Lewy condition (CFL) \citepads{1928MatAn.100...32C}. The divergence-free condition of the magnetic field, $\nabla \cdot \bm{B} = 0$, was numerically handled by the constrained transport method \citepads{2005JCoPh.205..509G}. A lower limit to the size of the time step was set to ensure the feasibility of the simulations.

The square grid has a physical domain spanning from 0 to 10 Mm, as shown in \cref{Fig: initial config}. The base resolution is 100$^2$. These cells have a width of 100 km. With adaptive mesh refinement (AMR), the effective resolution of the smallest cells can be increased to 3200$^2$ for six levels of AMR, which corresponds to a physical width of 3.125 km. The number of levels of AMR we used, and hence the smallest resolvable length scale, depends on the simulations itself and is mentioned when appropriate. Our highest physical resolution was the same as in \citetads{2020A&A...636A.112C}. A user-defined criterion based on the density was used to refine and coarsen the grid. 

In \cref{Sec: Dif cc} we compare the five cooling curves discussed in \cref{Sec: coolingcurves}. A summary of all the cooling curves available in the MPI-AMRVAC framework and their properties is given in \cref{App: ccs}. During a simulation, the lowest temperature of the adopted cooling curve is taken as the lowest possible temperature that the simulation can ever reach in order to avoid unphysical negative pressure values. An artificial Tfix parameter was set to ensure this. For the five cooling curves discussed above, this is not a problem because their lowest temperature is 1000 K at most and the lowest temperature in the simulations performed here does not reach to this value. However, in \cref{Sec: lowT} we discuss the effect of the low-temperature treatment of the cooling curves. Radiative cooling was handled as an additional source term. It was computed last to ensure that no other physical or numerical process could modify the temperature below this limit during a time step. The radiative energy loss was solved to use the exact integration scheme developed by \citetads{2009ApJS..181..391T}. For piece-wise cooling curves, the analytic formulae were used. However, for interpolatable tables, the temporal evolution function, denoted by $Y(T)$ in \citetads{2009ApJS..181..391T}, was calculated and tabulated in advance by numerically approximating the integral of the reciprocal of the cooling curve with respect to the temperature. This table was then used to calculate the energy loss by the optically thin radiative cooling cooling. More extensive details of the implementation for interpolatable tables are described in \citetads{2011CF.....42...44V}. The exact method is much faster than a traditional explicit scheme because it does not set a limit on the time step of the simulation. It is also more reliable than implicit schemes \citepads{2011CF.....42...44V}. The constant heating term, which is needed to initialise the setup with a background in thermal equilibrium, is given by \cref{Eq: heating}. It was handled as a user-set source term to the energy equation.

\section{Benchmark simulation}\label{Sec: benchmark}

The numerical setup is based on the work by \citetads{2020A&A...636A.112C}. Before the effect of several parameters on the formation of condensations can be compared, a benchmark needs to be set. We used the same cooling table and effective resolution as were used for the highest-resolution run in \citetads{2020A&A...636A.112C}, which are the SPEX\_DM table and 3200$^2$ . This resolution means that the smallest cells have a width of 3.125 km. We did not include thermal conduction, the effects of which are discussed later in this section. In addition to this difference, the largest difference is the orientation of the background magnetic field. In our case, the angle between the $x$ -axis and the magnetic field is 45$\degr$, while for the highest-resolution run in \citetads{2020A&A...636A.112C}, the angle was 54$\degr$. This small difference is significant, however, when in the benchmark simulation the magnetic field and combined perturbations of the two slow MHD modes travel in the same direction. This artificial symmetry ensures additional stability with respect to dynamical effects.

\begin{figure}[htbp]
  \resizebox{\hsize}{!}{\includegraphics{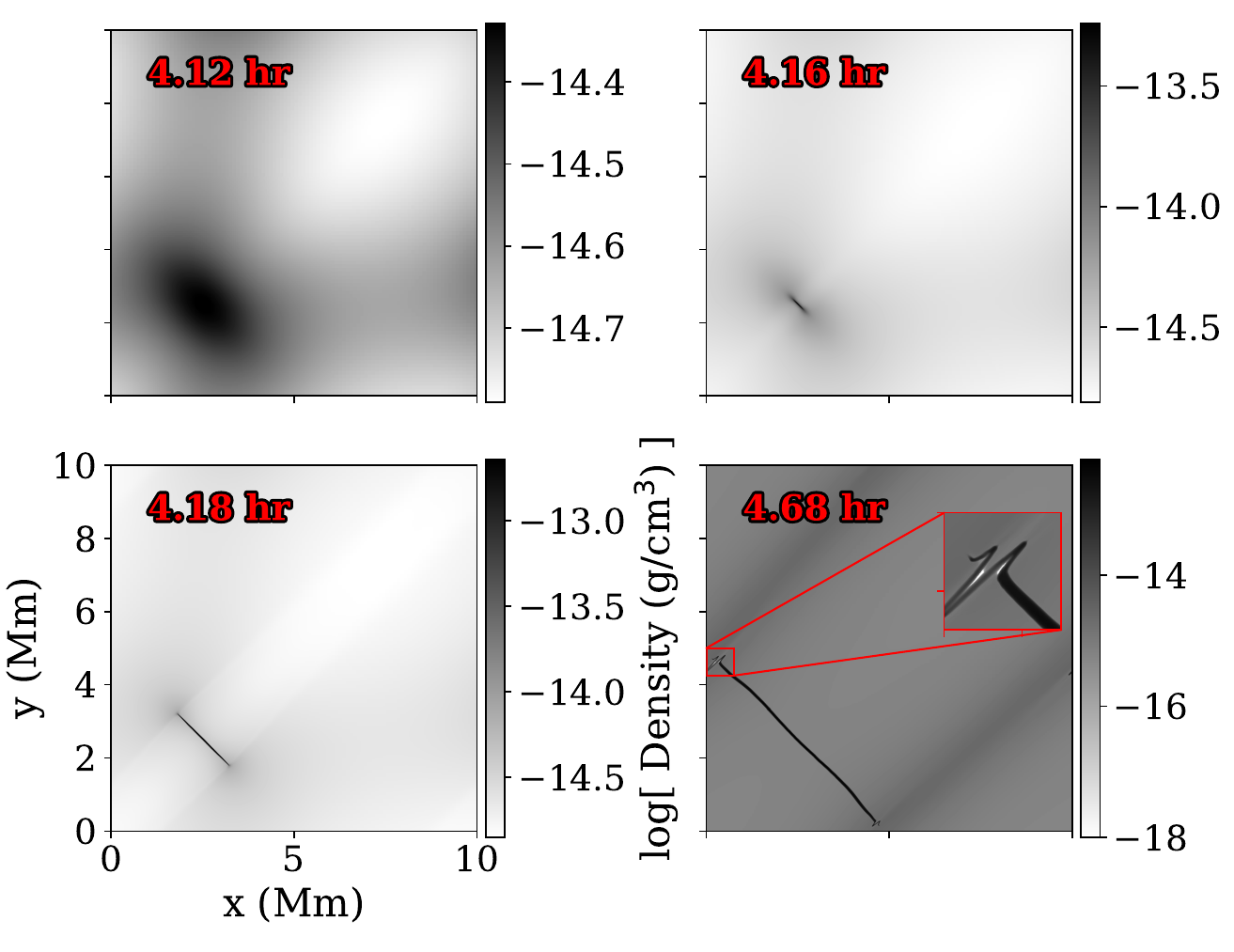}}
  \caption{Density view of the benchmark simulation using the SPEX\_DM table and a resolution of 3200$^2$. The top left and right panels have a density of twice and twenty times the background value, respectively. The bottom left panel has a density of 95 times the background value, and the bottom right panel denotes the end of the simulation due to the small time-step restriction.}
  \label{Fig: density_benchmark}
\end{figure}

The evolution of the thermal instability is shown in the four panels in \cref{Fig: density_benchmark}, where we plot the density view. The basic principles are the same as explained in detail in \citetads{2020A&A...636A.112C}. The top left panel is 4.12 hours after the initial state shown in \cref{Fig: initial config}. At this point, the interacting slow MHD waves are sufficiently damped due to the energy loss by radiative cooling. The thermal mode has started to grow because it is excited in an unstable regime. A region of increased density and decreased temperature starts to develop along the diagonal, that is, in the direction in which the interacting slow MHD waves were travelling. The lower temperature and hence thermal pressure causes an inflow of matter towards this spot due to a pressure gradient. The inflow occurs along the magnetic field lines because of flux freezing in the low plasma beta medium. After 4.12 hours, this region has reached a density of twice the initial background value. The increased density leads to a larger contribution of radiative cooling, which is dependent on the density squared. The cooler plasma also facilitates a higher cooling rate because the cooling curve globally increases from 10$^6$ to 10$^5$ K. This causes the catastrophic cooling associated with the thermal instability. In a mere 0.04 hours, or 2.4 minutes, the density increases to 20 times the initial background value, as we show in the top right panel, due to the counter-streaming inflows of matter along the magnetic field lines. This is the onset of the filament formation stage. Similar to \citetads{2020A&A...636A.112C}, we wish to note that with the term `filament', we mean a thin, high-density, low-temperature structure, not necessarily the solar feature related to prominences.

\begin{figure}[htbp]
  \resizebox{\hsize}{!}{\includegraphics{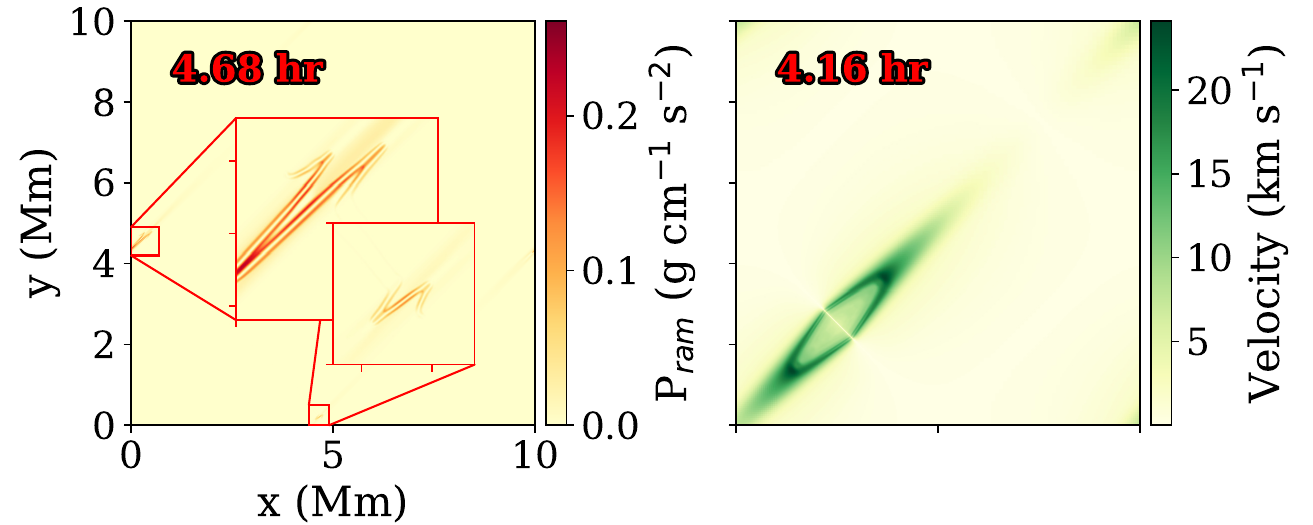}}
  \caption{\textit{Left}: Ram pressure view at the end of the benchmark simulation with insets zoomed in on the filament edges, where ram pressure differences fragment the filament. \textit{Right}: Total velocity magnitude after the filament has formed. It clearly shows the rebound shocks due to the formation of the filament.}
  \label{Fig: Pram_vel}
\end{figure}

Another 1.2 minutes later, a long, thin filament has formed, and its density has increased to approximately 95 times the background value. At this point, the filament has reached a temperature below 10\,000 K, at which the optically thin radiative cooling becomes less effective, see \cref{Fig: Coolingcurves}, ending the thermal instability. At the same time, a large part of the finite amount of mass in the strip perpendicular to the filament has been accumulated to the filament, limiting further growth in density. However, it should be noted that for the initial configuration typical for the solar corona, this density ratio matches the difference of two orders of magnitude that is observed between the solar corona and cool condensed prominences \citepads{2014LRSP...11....1P}. The rapid increase in density is due to accretion of mass along the magnetic field lines from plasma that is sucked into the central region. In this process, rebound shocks are observed to form that move outwards against the infalling material, as discussed in \citetads{2020A&A...636A.112C}. The depletion in the region perpendicular to the filament is apparent in the bottom left panel of \cref{Fig: density_benchmark}. In the right panel of \cref{Fig: Pram_vel}, the rebound shocks are shown in velocity view. As previously discussed in \citetads{2020A&A...636A.112C}, these rebound shocks are formed by the collision of the counter-streaming inflow of matter along the field lines that created the high-density filament. Matter is fed to the filament from the region in between the filament and the shock. The uniform medium sound speed and the Alfv\'en speed are approximately 150 and 583 km s$^{-1}$, which causes the velocities to become subsonic in the simulated reference frame. Similar rebound shocks have been observed in numerical simulations of prominence formation \citepads{2012ApJ...748L..26X} and coronal rain \citepads{2015ApJ...807..142F}. \citetads{2020A&A...636A.112C} quantified the nature of these shocks as slow MHD shocks.

From this point in our simulation, noticeable differences with the simulation of \citetads{2020A&A...636A.112C} start to arise. Owing to the lack of inclination between the magnetic field and the direction of the travelling waves, the filament has formed completely perpendicular to the magnetic field. The ram pressures at the front and backside of the filament are nearly equal, making it dynamically stable with respect to rotation. In the simulation by \citetads{2020A&A...636A.112C}, the misalignment allowed for a ram pressure imbalance, which in turn caused a pinching effect. This caused the filament to stretch, effectively thinning it and making it more susceptible to small velocity and pressure imbalances. In our simulation there is no torque, and the filament remains straight.

The fourth panel of \cref{Fig: density_benchmark}, after another half hour, shows the filament at its full extent with an inset showing fragmentation. A direct comparison of the timescale with the result of \citetads{2020A&A...636A.112C} is not possible because they used a different equilibrium temperature. At this point, the filament fragments. Unlike in \citetads{2020A&A...636A.112C}, where the filament was first twisted and thinned, the filament does not fragment everywhere, but only at the edges where it is thin enough. Although the location of the fragmentation is not the same as in \citetads{2020A&A...636A.112C}, the mechanism is. This is caused by small pressure or velocity imbalances, where small seeds may be numerical in origin (round-off errors). These imbalances grow due to the thin-shell instability. The best-known thin-shell instabilities are the linear \citepads{1983ApJ...274..152V} and non-linear thin-shell instability \citepads{1994ApJ...428..186V}. The first occurs when a thin shell is caught between ram pressure and thermal pressure, and in the second case, the thin shell experiences ram pressure on both sides. In the left panel of \cref{Fig: Pram_vel} we show the ram pressure view with insets zoomed in on the filament edges of this final moment. The large ram pressure differences around the horizontal fragmenting strands of lower density is apparent. This type of fragmentation of thin shells and slabs has also been observed in simulations of star formation in molecular clouds \citepads[e.g.][]{2003NewA....8..295H} and in the interaction of stellar winds with circumstellar disks \citepads[e.g.][]{2011A&A...527A...3V,2012A&A...547A...3V}. The fragmentation in our simulation shows that the pinching effect described by \citetads{2020A&A...636A.112C} is not a necessary condition, but it helps by thinning the filament.

Within the fourth panel of \cref{Fig: density_benchmark},  the inset shows a fragmenting edge with three perpendicular strands. This is the last frame of the evolution before our settings of the code, which warns the user when the CFL limit becomes too computationally demanding (time steps below 10$^{-9}$ in normalised units) and causes the simulation to stop. This small time step is caused by the significant decrease in density in between the fragmenting strands, which in turn reduces the CFL condition. When no artificial measures are taken to handle near vacuum conditions, this is the farthest into the non-linear evolution we can reach. In \cref{Sec: Extending} we explain how some bootstrap measures for low density can avoid this problem, and the simulation can continue.

As mentioned in \cref{Subsec: goveq}, we only included optically thin radiative cooling and a constant heating term. This also means that we did not account for thermal conduction, in contrast to the simulations by \citetads{2020A&A...636A.112C}. The effect of thermal conduction on a setup with optically thin radiative cooling, such as this thermal instability setup, is to facilitate a cut-off wavelength called the Field length. Perturbations with wavelengths shorter than this cut-off wavelength are stabilised, that is, they become damped because thermal conduction smooths these perturbations out. This was first described by \citetads{1965ApJ...142..531F} and was recently studied in detail by \citetads{2020A&A...636A.112C}. The Field length can be estimated as
\begin{equation} 
    \lambda_F = \sqrt{\frac{\kappa{(T)T}}{n^2\Lambda(T)}}, 
\end{equation}
\noindent where $n$ is the number density, $T$ is the temperature, and $\Lambda(T)$ and $\kappa(T)$ are the optically thin radiative loss and conductivity at a temperature $T$, respectively  \citepads{1965ApJ...142..531F,1990ApJ...358..375B}. This equation is not normalised. For a plasma under solar conditions, we can use the Spitzer conductivity parallel to the magnetic field \citepads{Spitzer2006}, defined as
\begin{equation}
    \kappa_{\parallel} = 8 \cross 10^{-7}~T^{5/2}~\text{erg cm$^{-1}$ s$^{-1}$ K$^{-1}$}.
\end{equation}
For the initial state and SPEX\_DM cooling curve, as the benchmark simulation, the Field length is about 10$^{10}$ cm and thus an order of magnitude larger than our perturbation wavelength. Hence the thermal mode will not be excited if we were to include thermal conduction in our setup. After the slow MHD modes are damped, the medium remains in balance. The idealised setup, without conduction, that we study here means that our Field length is zero and therefore leads to the scenario discussed in our benchmark case.

Thermal conduction also causes evaporation of cool, small parcels of plasma \citepads{1990ApJ...358..375B}. Hence, when thermal conduction is not included or is not propely resolved, it can lead to numerical errors and non-convergence, as discussed in \citetads{2004ApJ...602L..25K} and shown in \citetads{2010ApJ...720..652S}. The critical length scale that needs to be resolved is again the Field length because it is the characteristic length scale of the thermal conduction and depends on the local physical parameters of the plasma, on the thermal conductivity, and the optically thin radiative cooling rate. In setups without conduction, such as we discuss here, the local Field length is always and everywhere effectively zero. Numerical instabilities can thus arise at the size of the grid, that is, the smallest size in the simulation, because they are not damped out by conduction. Evaporation of cold plasma is thus not possible in our setups. Evaporation might alter the non-linear evolution when the filament has fragmented into treads, but this is not included in our work. However, the same issues arise when thermal conduction is present but the local Field length is unresolved, as would be the case if we were to incorporate thermal conduction. \citetads{2004ApJ...602L..25K} stated that at least three cells per local Field length are needed to obtain a physically correct solution. 

\begin{figure}[htbp]
  \resizebox{\hsize}{!}{\includegraphics{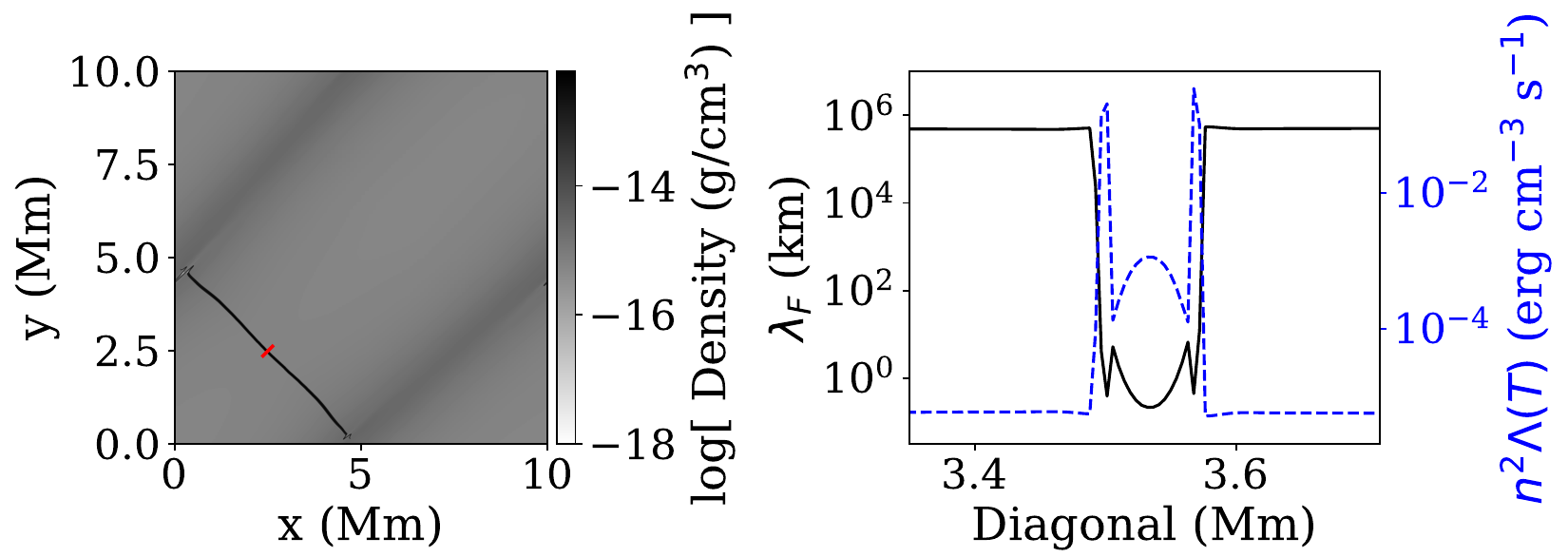}}
  \caption{Field length and optically thin radiative energy loss along a selected line. \textit{Left}: Density view at the end of the benchmark simulation with the selected line, along the diagonal of the grid and through the filament and transition region, superimposed in red. \textit{Right}: Field length and optically thin radiative energy loss along the selected line.}
  \label{Fig: fieldlength}
\end{figure}

A complementary view emphasises that this is because the transition region between the cold and dense condensation and the hot corona would otherwise not be correctly resolved, and hence thermal conduction would not be able to transport energy to the correct grid cells \citepads{2004ApJ...602L..25K}. A higher resolution minimises this problem. Nevertheless, even for our highest-resolution simulation, the local Field length, corresponding to the condensations, would still be unresolved by several orders of magnitude. At the typical temperatures and densities of condensations in the solar corona, the local Field length is of the order of a few metres, while high-resolution multi-dimensional simulations can resolve up to a few kilometres. In the right panel of \cref{Fig: fieldlength} we plot the Field length and energy loss by radiative cooling along a strip denoted by the red line in the left panel of the same figure at the end of the simulation. The selected strip is along the diagonal of the domain. It intersects the filament and thus contains information on the Field length in the filament and corona, as well as in the transition region between them. The Field length is several metres long in the filament and clearly unresolved. The Field length in the hot corona is well resolved. However, in the transition region, the Field length is even shorter than a few kilometres. This is due to the peak in radiative energy loss in the transition region. This peak is caused by the high cooling rates for temperatures between 10$^4$ and 10$^6$ K, in combination with a still high density with respect to the thin corona. The evaporation, driven by thermal conduction, in the transition region, that is, at intermediate temperatures, would not be resolved correctly. Even if we were able to resolve the local Field length everywhere in the domain along the magnetic field lines, the length scale of thermal conduction perpendicular to the field is still smaller because the perpendicular conductivity is 10$^{12}$ times lower \citepads{1965RvPP....1..205B}, and the associated resolution required would make a multi-dimensional MHD simulation computationally unfeasible. This limitation can be overcome if the conductivities are modified or isotropic conductivity is used, as discussed in \citetads{2010ApJ...720..652S}. The TRAC method that was recently developed by \citetads{2019ApJ...873L..22J} can patch this problem by artificially increasing the conductivity and decreasing the optically thin radiative cooling rate. This leads to a larger Field length and hence lifts the high-resolution restriction. The method was recently implemented in MPI-AMRVAC by \citetads{2021A&A...648A..29Z}, where it was extended to multi-dimensional MHD setups with anisotropic thermal conduction for the first time. However, this patch is typically used for the chromosphere-corona transition region and not for the transition region around in situ condensations.

\section{Effect of numerical settings}\label{Sec: effectNum}

In this section we investigate the effect of certain numerical, that is, unphysical, parameters or treatments of the evolution of the condensation formed by the thermal instability. First, we varied the resolution to understand differences between low- and high-resolution simulations. Second, we examined the low-temperature treatment of cooling curves. As mentioned in the preceding sections, there are several ways to handle the optically thin radiative cooling below 20\,000 K. At this temperature, the plasma becomes opaque and the optically thin assumption breaks down. Some cooling curves stop at this temperature, leaving the simulation without information. We used a setup similar to the benchmark setup and highlight the differences.

\subsection{Resolution}\label{Sec: Dif res}

The benchmark simulation was run with six levels of AMR, leading to an effective resolution of 3200$^2$. The width of the smallest cells was 3.125 km. Density blobs had a width of several cells, corresponding to a few tens to 100 km. This is of the order of the observational limit of the finest instruments, such as CRisp Imaging Spectro-Polarimeter (CRISP) \citepads{2008ApJ...689L..69S}. \citetads{2014ApJ...797...36S} showed statistically that the fine structure in post-flare loops might have a cross-sectional width smaller than 100 km. However, commonly used instruments, such as the Atmospheric Imaging Assembly on board the Solar Dynamics Observatory (SDO/AIA) \citepads{2012SoPh..275...17L} and the EUV Imaging Spectrometer (EIS) on board Hinode \citepads{2007SoPh..243...19C,2007SoPh..243....3K}, can only resolve down to around 1000 km \citepads{2006ApOpt..45.8689L,SDO_AIA_specs}. 
It is worthwhile to investigate the effect of the resolutions on the simulations to determine how much of the small-scale dynamics of the condensation formation process will be revealed at a given resolution. This is particularly timely, with upcoming possibilities to reach several dozen kilometre resolution using the Daniel K. Inouye Solar Telescope (DKIST) \citepads{2020SoPh..295..172R}.

We used the SPEX\_DM cooling curve and kept the base resolution constant to 100$^2$. The levels of AMR were varied from one to six. The MPI-AMRVAC framework has a block-based refinement scheme with a fixed refinement ratio between grid levels of two \citepads{2012JCoPh.231..718K}. When refining in 2D, a single block becomes four blocks, which each have a physical dimension half of that of the original block. We denote the simulations by the effective resolution. We thus examine the cases 100$^2$, 200$^2$, 400$^2$, 800$^2$, 1600$^2$ , and 3200$^2$. They correspond to a smallest width of 100, 50, 25, 12.5, 6.25, and 3.125 km, respectively. 

\begin{figure}[htbp]
  \resizebox{\hsize}{!}{\includegraphics{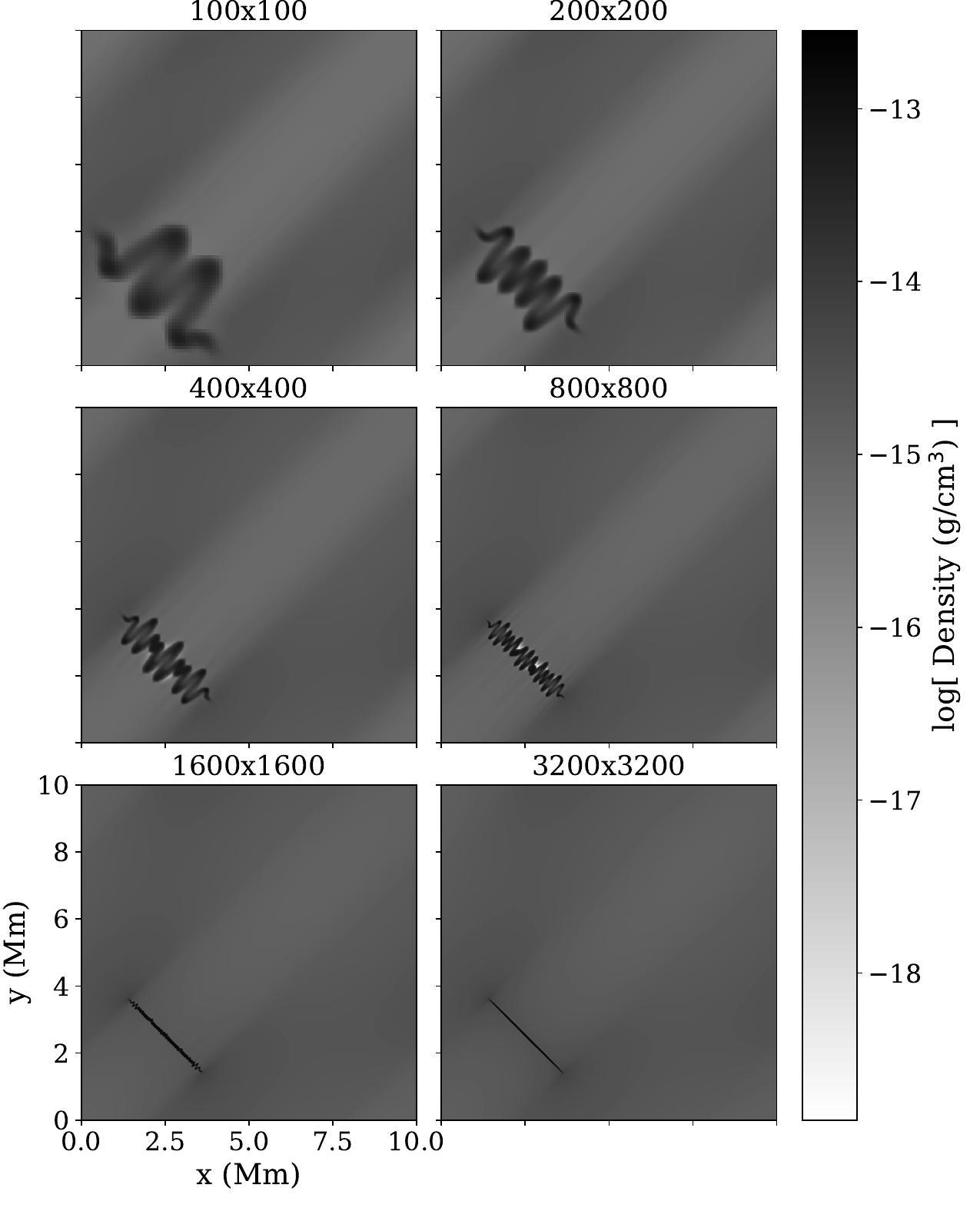}}
  \caption{Density views of the simulations at different resolutions. The snapshots are all taken after 4.22 hours.}
  \label{Fig: Dif_res_3x2}
\end{figure}

In the panels of \cref{Fig: Dif_res_3x2} we show the distribution of density after 4.22 hours. This is during the fragmentation phase. We chose this moment for the comparison because at this time, the 800$^2$ run encountered the small time-step limit. For all resolutions, a filament is formed, and the filaments are of approximately the same length. However, the width is different. The filament is thinner for higher resolutions. This behaviour has also been observed in simulations of filamentary structure in galaxy clusters by \citetads{2010ApJ...720..652S}. They discuss the need of resolving the Field length in every direction to obtain converged results. As mentioned before, the Field length here is effectively zero because we did not incorporate thermal conduction, and it is thus unresolved. The large gradients in quantities such as temperature or density necessarily induce small numerical errors. Strictly speaking, the 3200$^2$ simulation does not converge, but the results do not differ much from a resolution of 1600$^2$ to 3200$^2$. At the highest resolution, the filament has 20 cells across the diagonal to ensure a smooth transition region. This resolution hence truly resolves the internal filament structure.

From the first panel to the last, it is clear that the fragmentation occurs in each case, but demonstrates a different dominant wavelength along the filament. For a lower resolution, the wavelength with which the filament fragments is longer. In the case of the 100$^2$ run, there are only three wavelengths. The 800$^2$ run, in contrast, has approximately 12 wavelengths. This is because the wavelength of the thin-shell instabilities depend on the thickness of the shell. The wavelength is shorter for thinner shells. This has also been observed in simulations of stellar winds \citepads{2009A&A...508..751S}. In the 3200$^2$ run, no fragmentation has occurred. However, from the benchmark, we know that it will start to break up at a later time and only at the edges, as shown in \cref{Fig: density_benchmark}.

For each of the runs, we tracked the evolution of the maximum density and minimum temperature within the entire grid. These values are important properties of the filament. The maximum density reflects the amount of mass that was able to accumulate before the fragmentation begins. The total filament can still accumulate mass during the fragmentation phase. The minimum temperature denotes the temporal evolution of the thermal instability, which causes the formation of the condensation in our simulations.

\begin{figure}[htbp]
  \resizebox{\hsize}{!}{\includegraphics{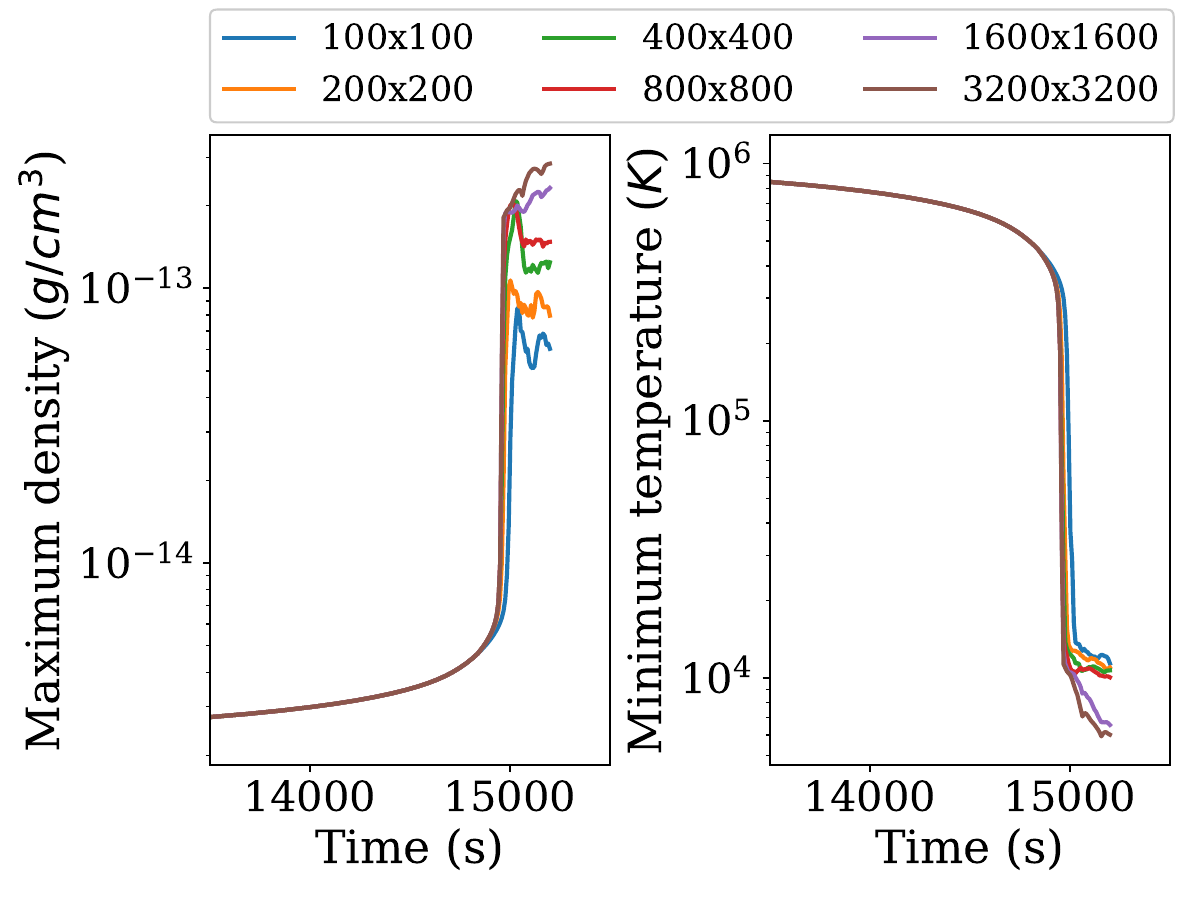}}
  \caption{Evolution of the maximum density and minimum temperature for the simulations at different resolutions. \textit{Left}: Maximum density. \textit{Right}: Minimum temperature.}
  \label{Fig: min_max_dif_res}
\end{figure}

In the left panel of \cref{Fig: min_max_dif_res} we show the evolution of the maximum density. It only shows the later phases because the density changes slowly while the slow waves are being damped. Several things can be observed. First, the simulations with lowest resolution, 100$^2$ and 200$^2$, do not reach the same maximum density as the other simulations at the end of the non-linear cooling phase. This can be explained by the fact that the filaments of these simulations fragment earlier and in larger strands. Second, the maximum density continues to increase for the highest resolution. This is because it has not yet started to fragment. Lastly, even though the filaments of the simulations with lower resolution all fragment, their evolution in terms of the instantaneous maximum density is still slightly different. This is due to the erratic behaviour of the fragmentation process, which is seeded from the numerical round-off errors, as mentioned earlier. 

The minimum temperature within the grid is shown in the right panel of \cref{Fig: min_max_dif_res}. The typical evolution of the thermal instability is very clear, where one can compare this panel to the left panel of \cref{Fig: Timescales}. For the four simulations with the lowest resolution, the minimum temperature reaches a plateau. During the fragmentation, the density has dropped or stopped increasing. This causes the cooling by optically thin radiation to be less efficient because it depends on the density squared. This does not occur in the simulations at the highest two resolutions. For these, the minimum temperature continues to decrease, although more slowly than during the non-linear cooling phase. This can be explained by the increase in density and the non-zero cooling rate, despite the decrease in cooling rate for temperatures below 10\,000 K for the SPEX\_DM table.

The density of the filament increases due to the accretion of plasma along the magnetic field lines, as discussed previously. Two other interesting characteristics of the filament are its surface area and accumulated mass. The simulations have only two dimensions. The mass is the surface mass, defined as the density of the filament times its surface area. The grid has a non-uniform resolution due to the adaptive mesh refinement; the region around the high-density filament has a higher resolution due to the refinement criterion. The surface area and surface mass were determined after extrapolating the grid to the highest resolution within the simulation. We defined the filament as the cells with a temperature lower than or equal to 15\,000 K. There are two reasons for this temperature limit. First of all, this temperature is near the end temperature reached by condensations formed by the non-linear phase of the thermal instability. Second, observations have shown that prominences and coronal rain formed in the solar corona reach these kinds of temperatures \citepads[see e.g.]{2014LRSP...11....1P,2020PPCF...62a4016A}.

\begin{figure}[htbp]
  \resizebox{\hsize}{!}{\includegraphics{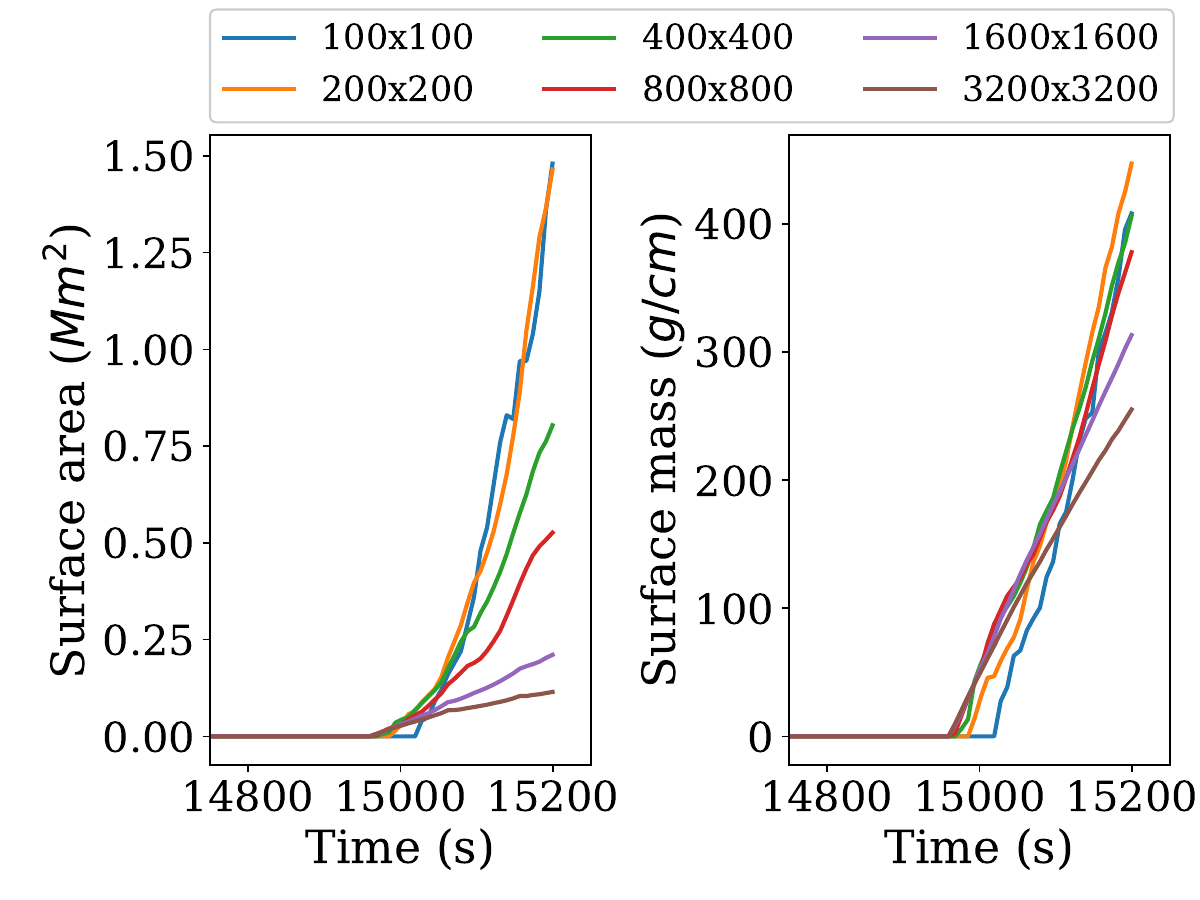}}
  \caption{Evolution of the surface area and surface mass of the filament for the simulations at different resolutions. \textit{Left}: Surface area. \textit{Right}: Surface mass.}
  \label{Fig: sa_sm_dif_res}
\end{figure}

The left panel of \cref{Fig: sa_sm_dif_res} clearly show that the surface area of the filament is smaller for simulations at higher resolutions. The reason is that at higher resolutions, thinner structures are formed. This behaviour was also present in the density views of \cref{Fig: Dif_res_3x2}. In the right panel of \cref{Fig: sa_sm_dif_res} we show the evolution of the surface mass. A lower resolution leads to more mass in the filament. This is likely related to the larger surface area that is covered by their filament. For these panels the exact values do not really matter because the final mass and surface are not reached due to the small time-step limit. For completeness, the total surface area and mass in the grid are 100~Mm$^2$ and 2340~g~cm$^{-1}$. In all cases only a small fraction of the total surface area and mass are accumulated in the filament, while the conservative discretisation employed and the use of double periodic boundary conditions ensures that the total mass in the domain is perfectly conserved.

\subsection{Low-temperature treatment}\label{Sec: lowT}

At low temperatures and high densities, such as for the condensations formed in these types of simulations, the plasma has a high opacity. It cannot cool efficiently and is optically thick. The plasma is not fully ionised for temperatures below 20\,000 K. Hence, self-absorption and photoionisation can occur within an optically thick plasma. This violates the assumptions of the collisional ionisation equilibrium and hence the optically thin radiative cooling \citepads{2003adu..book.....D}.

To correctly determine the energy loss by radiation, we would need to solve the full radiative transfer equations. However, this quickly becomes prohibitive in combination with multidimensional MHD, creating a need for more pragmatic measures that approximate the actual physics and keep simulations running. 

As an example, we considered the table of \citetads{2008ApJ...689..585C}. In the first low-temperature treatment, no changes to the table were made, but the Tfix parameter was used. This has the effect that when the plasma reaches the lowest temperature of the table, 11482 K for the Colgan table, instead of losing more energy, the plasma is kept at that temperature. Energy is artificially increased in the simulation. This is in contrast to the simulations without such a fix, where energy conservation is always in accord with the governing equation.

\begin{figure}[htbp]
  \resizebox{\hsize}{!}{\includegraphics{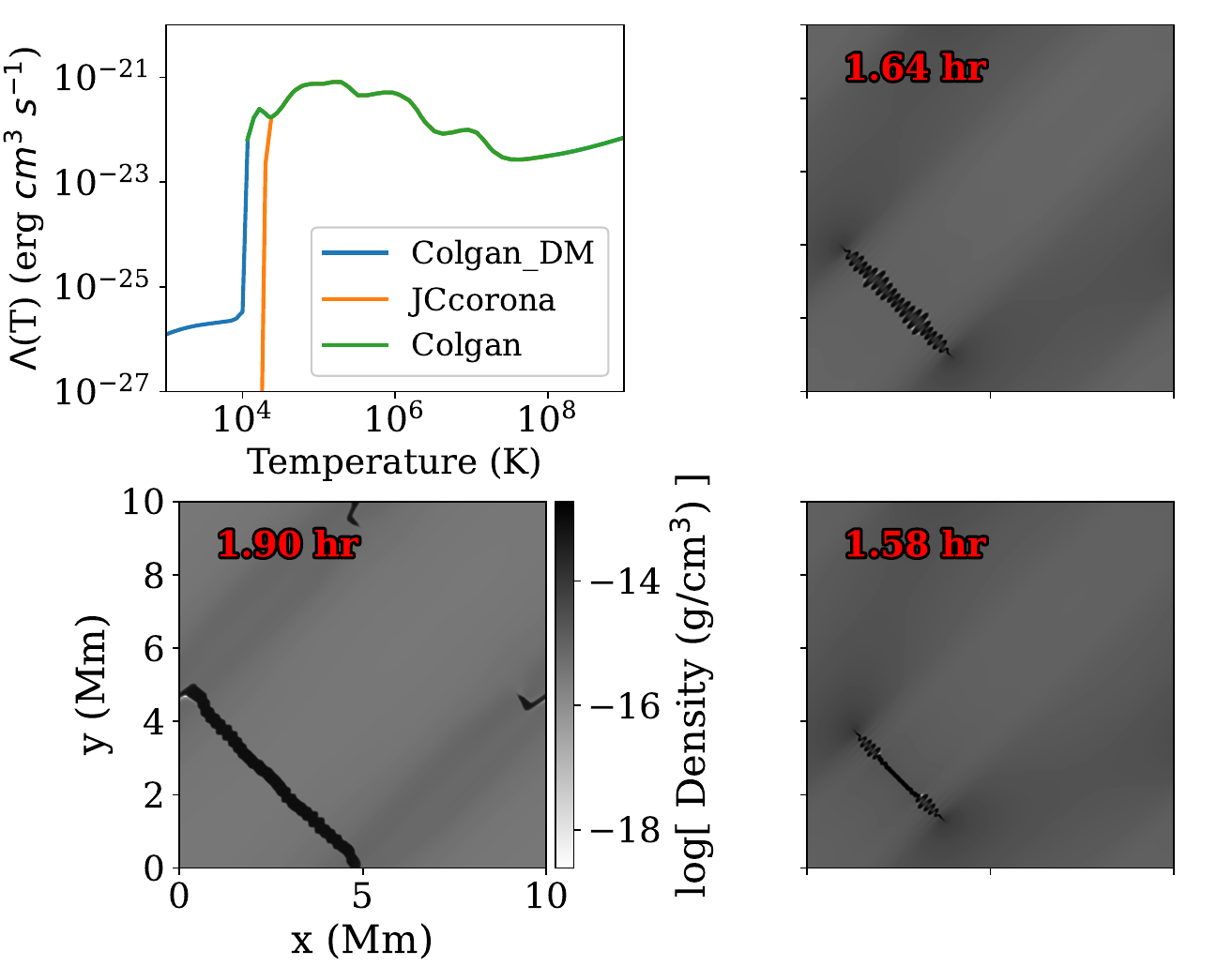}}
  \caption{Three cooling curves with different low-temperature treatment and their density view at the end of the evolution. \textit{Top left}: Cooling curves. \textit{Top right to bottom left and bottom right}: Density views at the end of the evolution. They correspond to the Colgan\_DM, JCcorona, and Colgan cooling tables, respectively.}
  \label{Fig: density_lowTbehaviour}
\end{figure}

The second method is commonly used in the astrophysics community. The cooling curve is modified to vanish for low, approximately 20\,000 K, temperature. The modified version of the Colgan table is the JCcorona table \citepads{2011ApJ...737...27X}, and it is shown in the top left panel of \cref{Fig: density_lowTbehaviour} together with the cooling curves of the other two treatments. This figure shows that by vanishing at 20\,000 K, a clear bump in the cooling rate is neglected. This bump is due to the Lyman-$\alpha$ radiation \citepads{1969ApJ...157.1157C}. Because the plasma is not fully ionised, it has been suggested in the past that neglecting the Lyman-$\alpha$ radiation is better than the optically thin radiative approximation at these temperatures \citepads{1983ApJ...265..497M}. Nevertheless, corrections for the radiative loss rate for Lyman-$\alpha$ have been attempted \citepads{1986ApJ...308..975A}. 

The third low-temperature treatment is also a modification, or rather an extension. It is the Colgan\_DM table, as discussed in \cref{Sec: coolingcurves}. It consists of a high-temperature part of the Colgan table \citetads{2008ApJ...689..585C}, augmented with cooling rates for temperatures below 10\,000 K by \citetads{1972ARA&A..10..375D}. A low-temperature extension is justified because even though the optically thin approximation fails, there will still be some radiative cooling. The cooling is due to collisional excitation of singly charged ions of O, C, N, Si, Fe, Ne, and S with thermal electrons and neutral hydrogen. There is also a contribution by collisions between neutral hydrogen with electrons, as described in \citetads{2009A&A...508..751S}. The cooling rate at these low temperatures is several orders of magnitudes lower than at high temperatures, as shown in the top left panel of \cref{Fig: density_lowTbehaviour}, but is never actually zero (as adopted without this extension of the cooling curve).

To show the effect of the low-temperature treatment, we ran the thermal instability setup for the three cooling curves discussed in the previous paragraphs. We used four levels of AMR, leading to an effective resolution of 800$^2$ and a width of the smallest cell of 12.5 km. From the previous subsection we have learned that this will yield fragmentation at larger wavelengths than for higher resolutions. We would like to emphasise that we studied a single cooling curve here, the Colgan table, and applied different treatments for the low-temperature regime. In \cref{Sec: coolingcurves} we discuss the comparison between different cooling curves that might have the same low-temperature treatment, such as the Colgan\_DM and SPEX\_DM tables.

The density views at the end of the simulations, the moment at which the low density limits our time step, are shown in the right and bottom panels of \cref{Fig: density_lowTbehaviour}. The top right, bottom left, and bottom right panels correspond to the Colgan\_DM, JCcorona, and Colgan cooling tables, respectively. In all cases a filament is formed, but there are differences in morphology. For the Colgan and Colgan\_DM  the filaments fragment, but for the JCcorona table, it becomes thicker and only the edges stretch. The difference with and without the \_DM part is not so pronounced. In the case of the Colgan table, the fragmentation occurs in a similar way as for the Colgan\_DM, but the filament remains unperturbed in the centre. We are led to conclude that the vanishing and the ceasing of the optically thin cooling undermines or at least delays the thin-shell instability. Another way of viewing this is that the radiative cooling becomes ineffective for the JCcorona table. The evolution of the filament is therefore governed by its dynamics, which appear to be near equilibrium.

\begin{figure}[htbp]
  \resizebox{\hsize}{!}{\includegraphics{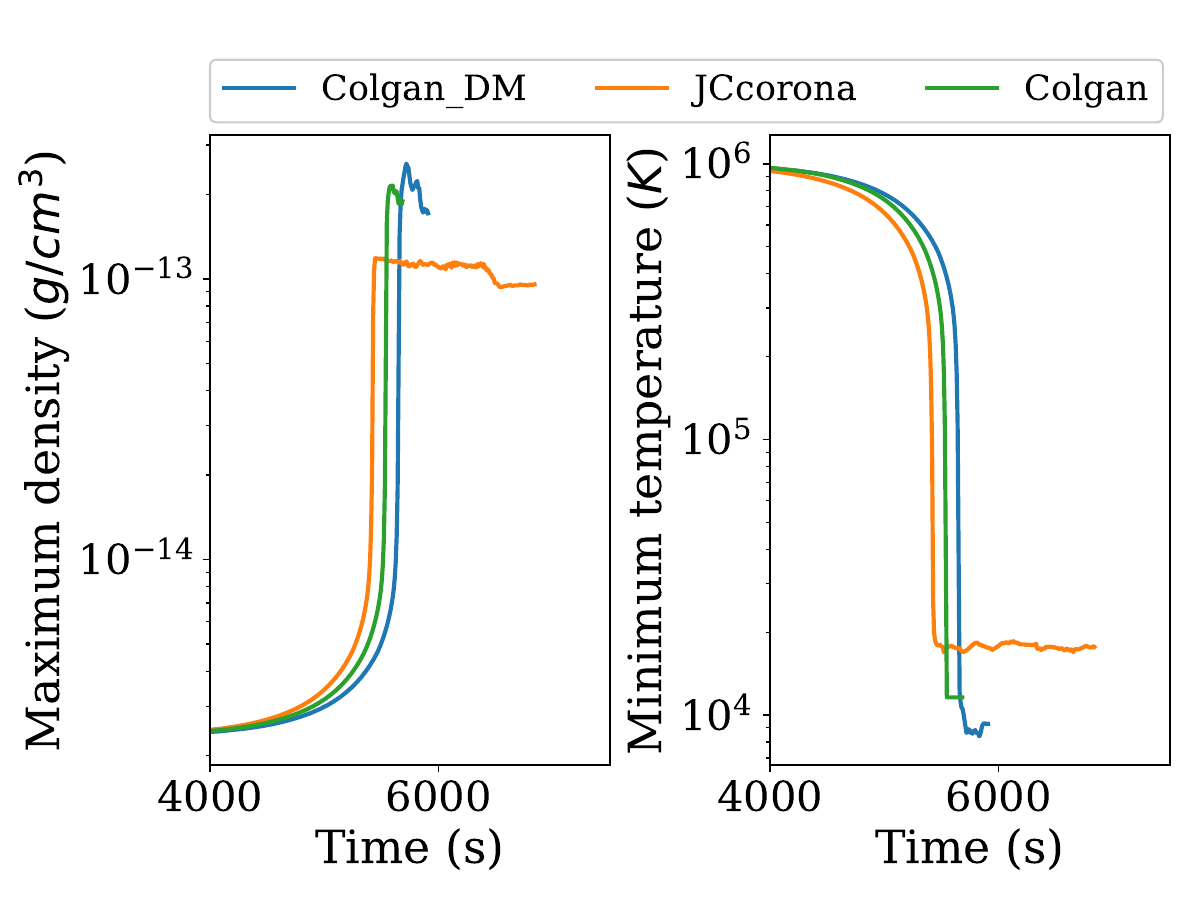}}
  \caption{Evolution of the maximum density and minimum temperature for the simulations with the Colgan\_DM, JCcorona, and Colgan cooling tables, which only differ in their low-temperature treatment. \textit{Left}: Maximum density. \textit{Right}: Minimum temperature.}
  \label{Fig: min_max_lowT}
\end{figure}

As before, we tracked the evolution of the maximum density and minimum temperature in the grid for the three simulations. The results are presented in \cref{Fig: min_max_lowT}. We note a small difference in timescale of the thermal instability, although the cooling rate at 10$^6$ K is exactly the same. Due to interpolation when constructing the interpolatable cooling curves numerically, very small differences in the cooling curves arise. This can lead to slightly more or less energy loss after many iterations. Hence the timescale of the thermal instability might be slightly longer or shorter, respectively. In all the cases the instability begins and a filament is created, as shown in the density views of \cref{Fig: density_lowTbehaviour}. 

In the right panel of \cref{Fig: min_max_lowT} the different low temperature treatments are clear. The temperature remains constant for the Colgan table after cessation of the thermal instability. For the simulation with the JCcorona table, the thermal instability stops abruptly at 20\,000 K, and its further temperature evolution is determined by the changes in pressure and density. For the Colgan\_DM table, the evolution reaches lower temperatures and stops because of the fragmentation and small time-step limit.

The evolution of the maximum density is also different, as expected from the density views. This is shown in the left panel of \cref{Fig: min_max_lowT}. The results of the Colgan and Colgan\_DM table are quite similar, but the simulation with the JCcorona table reaches a much lower maximum value. It also remains at this value afterwards because there is no apparent fragmentation to alter the density significantly. The density views also showed that the extent of the filament perpendicular to the magnetic field lines is different. This might be because the onset of fragmentation, which varies, appears to limit the perpendicular growth.

We can conclude that the low-temperature treatment has a strong effect on the density and temperature evolution of condensations. Indirectly, it has an effect on the fragmentation process, and hence on the morphology of condensations. In reality, this means that the true physical conditions of being non-optically thin when the condensation has formed should be accounted for in future simulations.

\section{Effect of changing the cooling curve}\label{Sec: Dif cc}

\begin{figure*}[t!]
\centering
   \includegraphics[width=17cm]{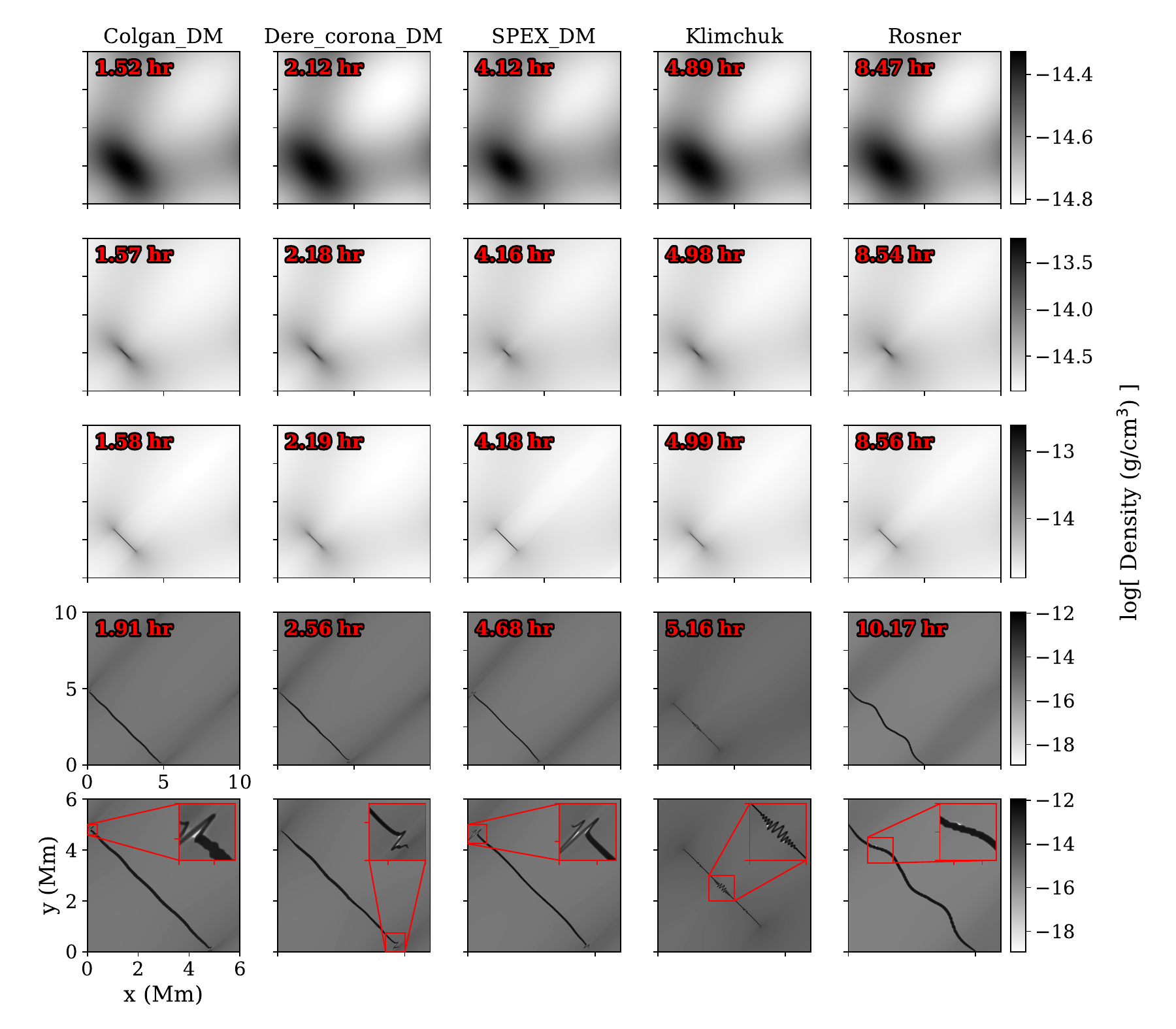}
     \caption{Density views during the simulations with different cooling curves. Each cooling curve has its own \textit{\textup{column}}. The \textit{\textup{rows}} represent similar moments and are scaled to the same density colour bar. The first three rows are when the filament has a maximum density of approximately 2, 20, and 95 times the background density. The fourth row is the last moment, at which the time step becomes restrictive. The last row is a zoom-in onto the filament of the last moment, it also contains a zoomed-in inset of the low-density region.}
     \label{Fig: Dif_CC_complete}
\end{figure*}

In the previous section, the effect of numerical settings, such as the resolution and low-temperature treatment of the radiative cooling, were investigated. Here we discuss the effect on MHD simulations of completely changing the optically thin radiative cooling curve. To this end, we again study the effect on the condensation process of the thermal instability. The general setup is discussed in \cref{Sec: benchmark}, with six levels of AMR corresponding to 3.125 km. We used the five cooling curves discussed in \cref{Sec: coolingcurves}. The Colgan\_DM, Dere\_corona\_DM, and SPEX\_DM are interpolated tables, which are extended for low temperatures with the DM table. The Klimchuk and Rosner tables are piece-wise power laws.

The evolution of the thermal instability and formation of a condensation for the five cooling curves is shown in the panels of \cref{Fig: Dif_CC_complete}. Each row denotes a different phase. The first four rows are the same four times as discussed for the benchmark simulation of \cref{Sec: benchmark}. The benchmark is the SPEX\_DM run. The last row shows a zoom-in of the filament and has an inset that is zoomed into the low-density region that triggers the small time-step condition.

For all five cooling curves, the first stages of the evolution are similar. The thermal instability is not obstructed by changing the cooling curve. A thin filament with a maximum density of 95 times the background value always forms. From the fourth and fifth row, it is clear that the final shape, that is, the morphology, can be different. The filaments of the simulations with the Colgan\_DM, Dere\_corona\_DM, and SPEX\_DM tables remain straight, with fragmentation only at the edges. They also have approximately the same length. The filament belonging to the Klimchuk table is much shorter and fragments at several places, but most prominently at the centre. The filament of the Rosner table is longer and becomes wavy. There is again a suggested link between the length of the filament and the amount or onset of fragmentation. The perpendicular growth appears to be hindered as soon as the filament fragments.

These global differences in morphology are due to the low-temperature treatment of the cooling curves, because for all cooling curves, the thermal instability creates a condensation. It is thus not surprising that the results for the first three cooling curves are similar. They have the same low-temperature treatment. There are still small differences between these three cooling curves because their earlier evolution was slightly different. This difference leads to small variations in density and pressure and hence in the fragmentation by the thin-shell instability.

Based on the results of \cref{Sec: lowT}, we can try to explain the behaviour of the other two curves. We have noted before that when the cooling curve vanishes, the filament did not fragment as easily. This is also the case for the Rosner table. Its cooling rate does not vanish completely at 20\,000 K, but drops off very fast and does not reach a low-temperature plateau such as for the tables with \_DM extension. The Klimchuk table is much coarser. Its cooling rate for low temperatures remains at least an order of magnitude higher than for the tables with the \_DM extension, see \cref{Fig: Coolingcurves}. A possible explanation could be that a higher cooling rate at low temperatures leads to small temperature and hence pressure variations or blobs within the filament. This in turn leads to easier fragmentation by thin-shell instabilities due to the presence of small seeds.

We also wish to note that by changing the cooling curve, we inevitably also modify the heating of a given model. To recall the heating function, given by \cref{Eq: heating}, it is constant, depends on the cooling curve, and was chosen this way to ensure thermal equilibrium of the initial state. Thus the differences observed here between models with different cooling curves are due to the combined effect of a different cooling rate and background heating.

\begin{figure}[htb]
  \resizebox{\hsize}{!}{\includegraphics{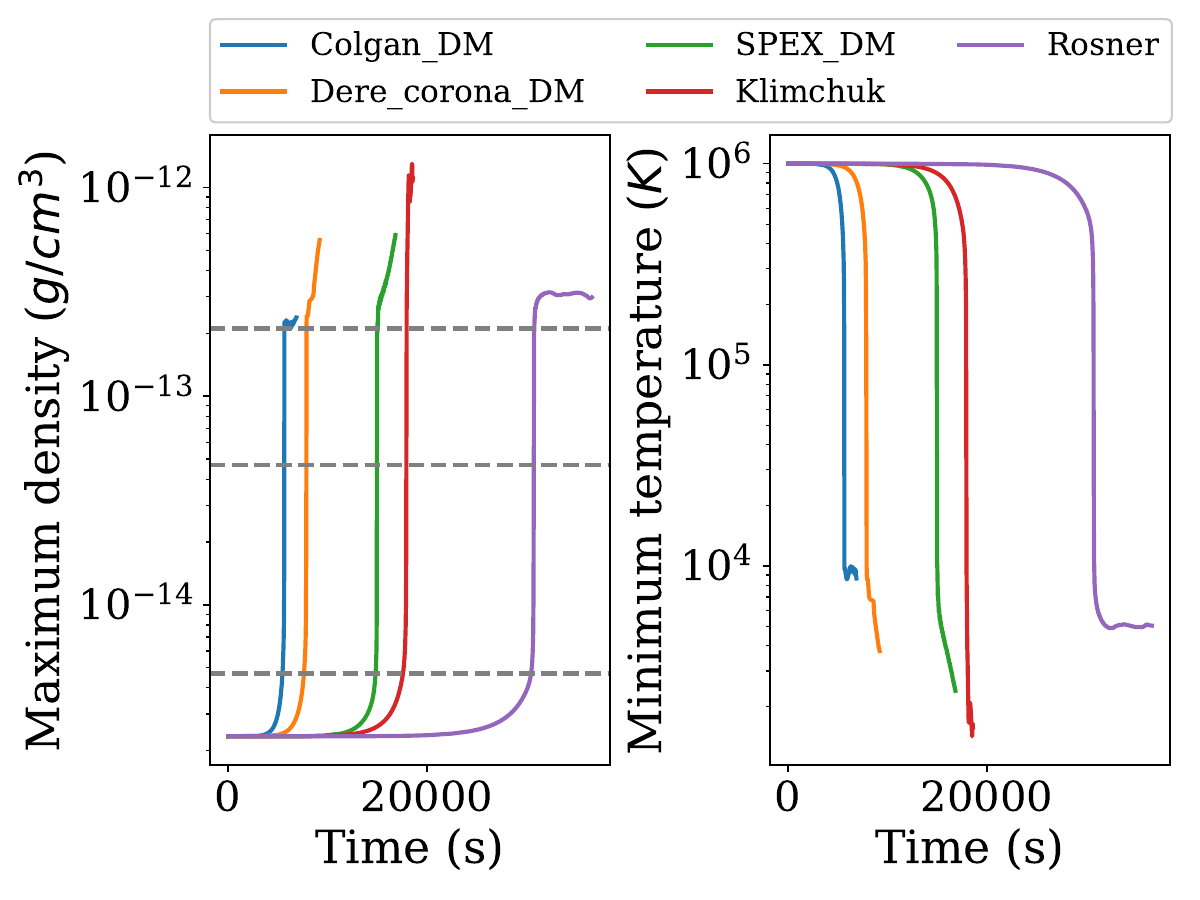}}
  \caption{Evolution of the maximum density and minimum temperature for the simulations with different cooling curves. \textit{Left}: Maximum density. The horizontal dotted grey lines denote the maximum densities of the rows of \cref{Fig: Dif_CC_complete}. \textit{Right}: Minimum temperature.}
  \label{Fig: min_max_dif_cc}
\end{figure}

The right panel of \cref{Fig: min_max_dif_cc} shows the evolution of the minimum temperature. As previously discussed, the linear timescales are very different on the basis of the cooling curve itself (see \cref{Fig: Timescales}). The numerical simulations indeed confirm this. The Rosner cooling curve has a much lower cooling rate at 10$^6$ K. The thermal instability takes much longer to form a condensation. In addition to the difference in timescales, it is important to note that the final temperature is different. However, all temperatures reach below 20\,000 K, at which optically thin cooling is no longer strictly valid. The maximum density reached is also different, as shown in the left panel of \cref{Fig: min_max_dif_cc}. These two effects are intertwined. The simulations stop because of the fragmentation, which is different even for cooling curves with the same low-temperature behaviour. Small differences in the cooling curves during the evolution of the filament can cause structural differences and hence small differences in the final result of certain parameters, although the global density view looks similar.

The Klimchuk and Rosner cooling curves give a totally different result. The minimum temperature of the Klimchuk simulation continues to decrease and causes its maximum density to increase. This is most likely because its cooling rate does not decrease much compared to the other cooling curves, and hence energy is still efficiently radiated away. This allows the temperature and pressure to drop and facilitates the accumulation of more matter. The simulation with the Rosner cooling curve reaches a more steady end phase, with both the maximum density and temperature quite stable for more than an hour. This behaviour can also be assigned to the low-temperature behaviour of its cooling curve, as discussed before.

It seems trivial that the timescales are different, but it has important consequences. The setup of the thermal instability we used is simplified. In the real world, many forces and effects occur at once, and many external perturbative events can take place. In the context of solar physics and prominence formation, a flare or wave might perturb the environment. If this occurs after a given time, then the condensation might have formed for one cooling curve, but not for another.

\begin{figure}[htbp]
  \resizebox{\hsize}{!}{\includegraphics{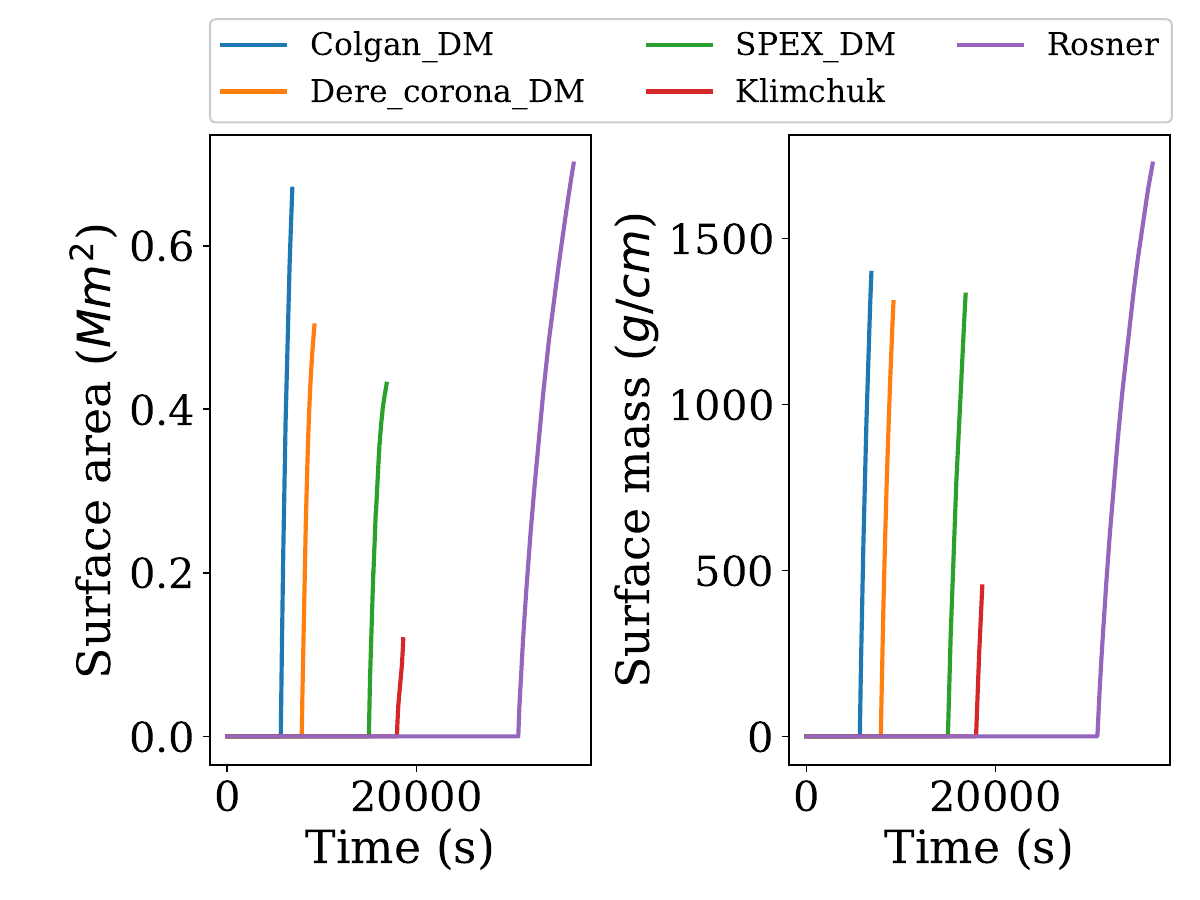}}
  \caption{Evolution of the surface area and surface mass of the filament for the simulations with different cooling curves. \textit{Left}: Surface area. \textit{Right}: Surface mass.}
  \label{Fig: sa_sm_dif_cc}
\end{figure}

For the simulations with varying resolution, we track the evolution of the surface area and surface mass of the filament, defined by the cells with a temperature lower than 15\,000 K. The results are shown in \cref{Fig: sa_sm_dif_cc}. The surface masses of the filaments of the three \_DM tables are nearly equal, but their surface areas are not. Just like for their maximum density, small variations in the physical variables and small differences in the fragmentation lead to differences in surface area. The SPEX\_DM table has the highest maximum density and the smallest surface area of the three curves. The opposite is true for the Colgan\_DM table. This is because the slightly earlier fragmentation in the case of the SPEX\_DM table leads to more contracted or compressed density clumps in the strands. The combination of the two effects leads to the equal surface mass.

For the Rosner table the filament did not fragment. It continued to grow in both directions, along and perpendicular to the magnetic field. This is clear from the evolution of the surface area and surface mass, and from the density views. The Klimchuk table produced distinctly different results. The surface area is at least four times smaller than for the other tables and twice the surface mass. Due to the strong fragmentation in this case, the growth is halted. Its perpendicular length is also much shorter than for the other curves, as shown in the bottom panels of \cref{Fig: Dif_CC_complete}. The smaller filament has thus a smaller surface area, and this leads to a lower surface mass. The difference in surface mass is smaller because the overall density of the filament is higher, counteracting the small surface area.

\section{Extending far into the non-linear regime}\label{Sec: Extending}

We can now use the insights gained in the previous sections to study the evolution of a condensation formed by the thermal instability far into it non-linear regime. To do this, we study the fragmentation phase in more detail. 

All the simulations we showed so far and the similar simulations performed by \citetads{2020A&A...636A.112C} were terminated due to a small time-step restriction. This small time step is caused by the Courant–Friedrichs–Lewy condition (CFL). This is a necessary condition for stability when any kind of explicit time-marching scheme is used for the essentially hyperbolic MHD equations that were first derived by \citetads{1928MatAn.100...32C}. For a given resolution normalised $\Delta x,$ it is given by
\begin{equation}
    C  = \frac{\abs{v}\Delta t}{\Delta x}, 
\end{equation}
\noindent where $C$ is the Courant number, $\Delta t$ is the normalised time step, and $v$ is the fastest velocity at which information travels through the grid. In this simulation, it is the local velocity magnitude plus the local fast magnetosonic speed. From a physical point of view, this means that the explicit time step has to be shorter than the time required for the fastest wave to propagate from one cell to another. More information can be found in standard textbooks on magnetohydrodynamics and numerical modelling \citepads[see e.g.][]{GoedbloedJP2019MoLa}. 

In the simulations presented here, the short time step was caused by the sudden drop in density in between the fragmented strands perpendicular to the filament and aligned with the magnetic field. For our simulations, the time step can be approximated in terms of the density and resolution as 
\begin{equation}
    \Delta t \approx 0.1 \Delta x \sqrt{\rho},
\end{equation}
\noindent where we used a Courant number of 0.8 and used the fast MHD wave velocity. The actual restriction on the time step thus comes from the fact that the filament collects almost all matter on an initial field line, and our periodic boundaries then mean that no additional mass is available to counter the drop in density.

In order to overcome this problem, we made use of a numerical bootstrap technique. This amounts to introducing an artificial lower boundary to the allowed density, that is, if the density drops below a certain value, we adjust it. This is one of the different uses of the low value fixes, such as negative pressure handling, within MPI-AMRVAC. The framework has several methods to handle low values, such as replacing the triggered cell by a user-set lower bound or by averaging in a box around the cell. The average method is normally more physical, but in our case, the low density occurs in between the fragmented strands. This causes the averaged value to be very high compared to the original low density. We therefore made use of the replace method. This method violates the conservation of mass in the system by introducing additional plasma. 

Because we aimed to extend a simulation for a long time, we needed a reasonable time step for a given iteration while still resolving the physical details. We chose to use four levels of AMR, that is, an effective resolution of 800$^2$. We used the SPEX\_DM table as for the benchmark case. Our lower density bound, also denoted by $\rho_{min}$, was set to 0.01 in normalised units, which is about 10$^{-17}$~g~cm$^{-3}$. This was sufficient to simulate the entire further non-linear evolution we discuss next.

\begin{figure}[htbp]
  \resizebox{\hsize}{!}{\includegraphics{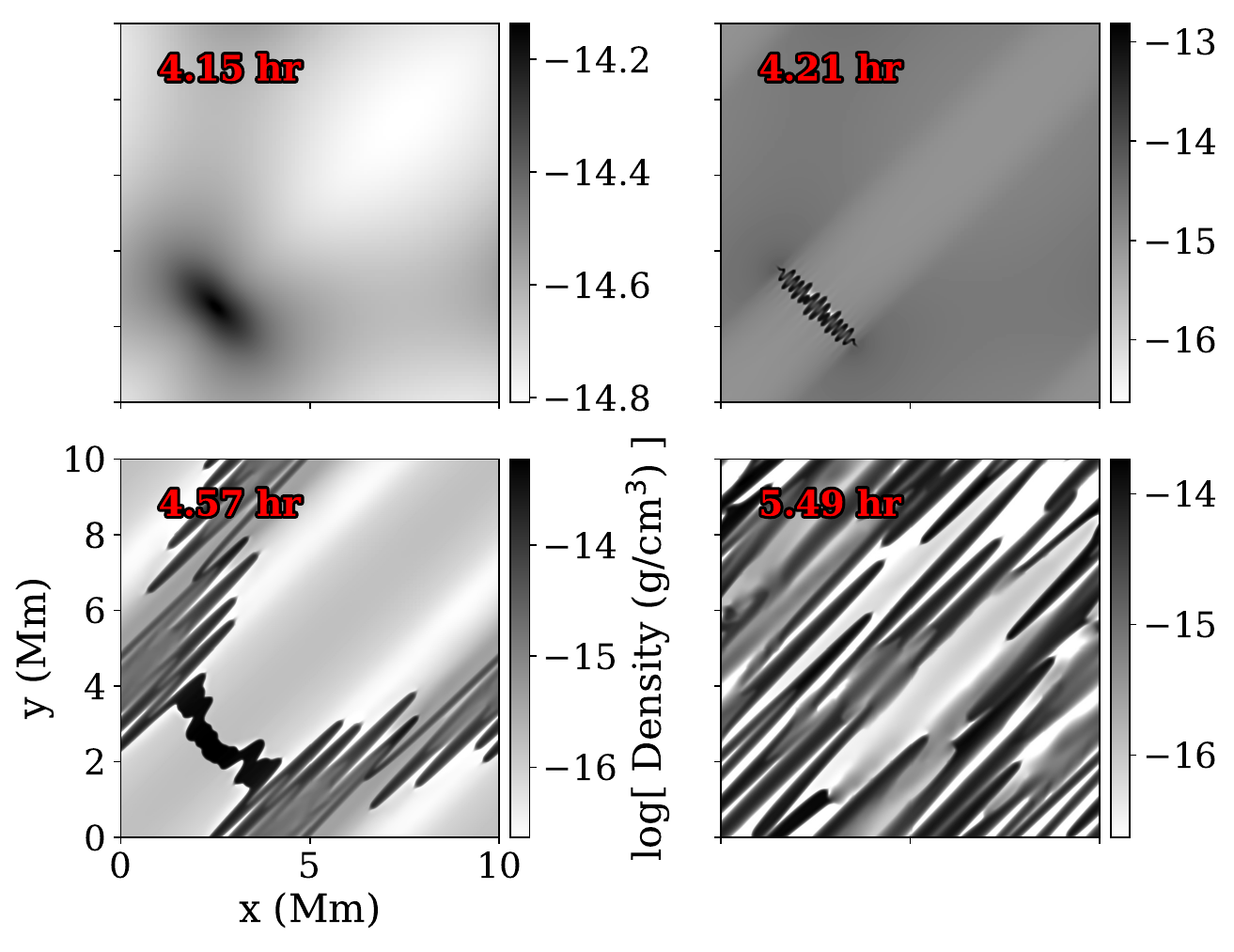}}
  \caption{Density view of the full evolution of the thermal instability. \textit{Top row}: Onset of the thermal instability and the moment in which the low-density bootstrap measure kicks in, in the left and right panel, respectively. \textit{Bottom left}: Moment in which 1\%\ of the area is at the prescribed lowest allowed density. \textit{Bottom right}: Last moment of the simulation. The animation of this simulation is available at \url{https://erc-prominent.github.io/team/jorishermans/}}
  \label{Fig: density_extended}
\end{figure}

In \cref{Fig: density_extended} we show the evolution of thermal instability and its fragmentation. In the top left panel we show the starting point. We started from the 800$^2$ simulation of \cref{Sec: Dif res} at the point where the condensation started to form by the thermal instability. The lower density bound is not yet reached here. In the top right panel we show the fragmenting filament at the moment in which the low-density treatment starts working. The white regions in between the strands have this low density value. The bottom left image depicts the density distribution roughly 22 minutes later. The highest density has decreased, but most of the mass in the filament has coagulated in the centre of the filament, while the outer parts become strands guided by the magnetic field. At this stage, 1\%\ of the total number of cells have been affected by the treatment. These are the white areas between the strands. In the bottom right panel we show the density distribution after one hour and twenty minutes. We ended the simulation at this point because about 5\%\ of additional mass was introduced to the simulation by the continuous bootstrap strategy. At this point, the filament is completely broken up and the strands are scattered throughout the grid. However, their orientation is still aligned with the magnetic field, which is to be expected because it is a low beta plasma with a strong magnetic field. There are extended regions with low density.

\begin{figure}[htbp]
  \resizebox{\hsize}{!}{\includegraphics{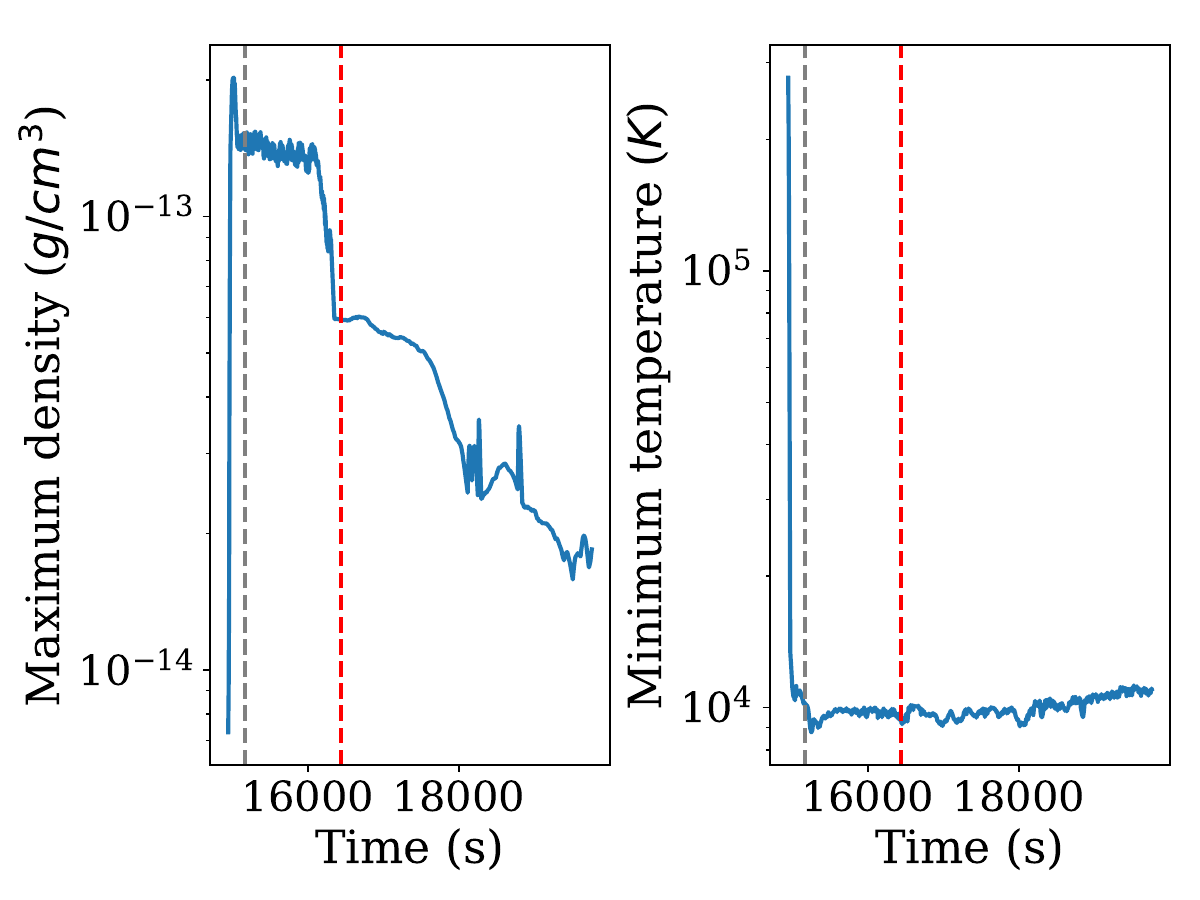}}
  \caption{Evolution of the maximum density and minimum temperature for the simulation of the full evolution of the thermal instability. \textit{Left}: Maximum density. \textit{Right}: Minimum temperature. The evolution is shown from the start of the simulation, that is, the onset of the thermal instability. The dotted grey and red lines correspond to the top right and bottom left density views of \cref{Fig: density_extended}, respectively.}
  \label{Fig: min_max_extended_run}
\end{figure}

In \cref{Fig: min_max_extended_run} we show the evolution of the maximum density and minimum temperature in the grid. As mentioned previously, the starting point is the onset of the thermal instability, which can be seen as the rise and drop in density and temperature, respectively. In these figures we can study the full evolution in contrast to the results in the previous sections.

The temperature evolution is straightforward to interpret. Because of the thermal instability, the temperature drops non-linearly. The drop stops at around 10\,000 K. This is to be expected because the optically thin radiative cooling becomes less efficient. During the fragmentation and redistribution of matter, the minimum temperature remains approximately constant. However, there appears to be a slight increase in temperature while the filament is redistributed. It is noted here, and further explained in our \cref{App: HD}, that the fragmentation and dynamics seen in this low plasma beta situation are very different from purely hydrodynamic settings.

The maximum density increases to its maximum value due to the pressure gradient. The gradient facilitates the flow of material towards the filament along the magnetic field lines. The accretion of matter is clearly shown in the top right panel of \cref{Fig: density_extended} by the lower-density regions perpendicular to the filament. Through collisions within the filament and fragmentation, the maximum density drops slightly. The grey line corresponds to the time in the top right panel of \cref{Fig: density_extended}, while the red line corresponds to the bottom left panel. At the latter time, the filament breaks up completely, which causes larger drops in maximum density. During the redistribution, several strands of plasma might interact with one another, causing small bumps in the maximum density, as shown in \cref{Fig: min_max_extended_run} at about 18\,000 to 18\,500 seconds. However, the global trend is a decrease in maximum density, meaning a redistribution of the material over the box along the magnetic field lines. This behaviour is quite reminiscent of coronal rain or downflows in the legs of prominences. Moreover, despite the initial orientation of the filament, which was perfectly orthogonal to the magnetic field, the final state we obtain is one of many filaments that are clearly aligned with the background field.

\begin{figure}[htbp]
  \resizebox{\hsize}{!}{\includegraphics{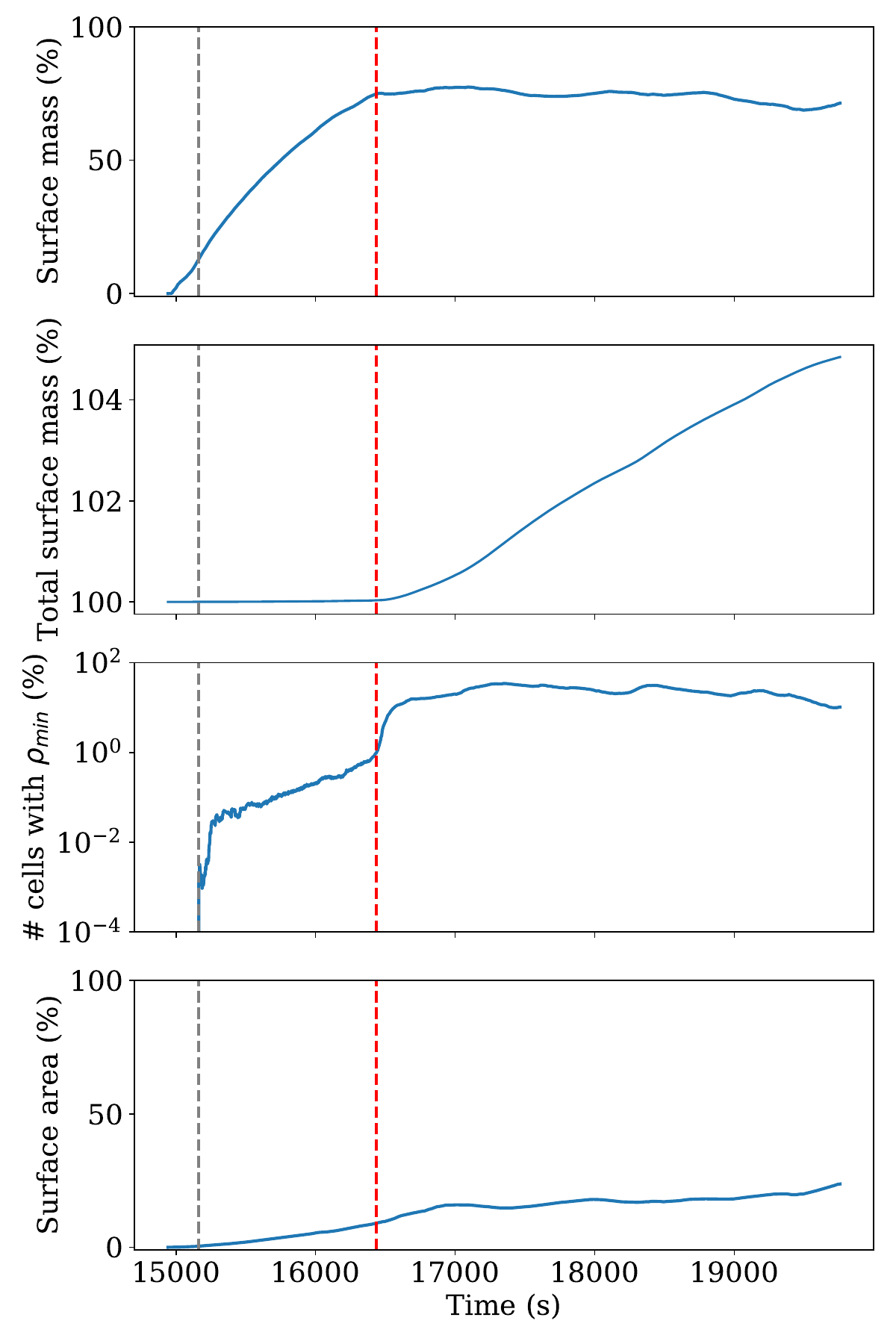}}
  \caption{\textit{Top panel}: Evolution of the surface mass of the filament. \textit{Second panel}: Evolution of the total surface mass in the grid. \textit{Third panel}: Evolution of the percentage of the number of cells or area with the prescribed lowest allowed density. \textit{Bottom panel}: Evolution of the surface area of the filament with respect to the total area.}
  \label{Fig: sa_sm_extended_run}
\end{figure}

To study this complete fragmentation phase, we used a numerical trick to prevent the time step from becoming impractically short. Nevertheless, we need to investigate the effect of this bootstrap approach on the physical behaviour of the numerical simulation. In \cref{Fig: sa_sm_extended_run} we show the evolution of the percentual surface mass in the filament, that is, temperature lower that 15\,000 K, the total surface mass, the number of cells or amount of surface area with the artificial limit value, $\rho_{\mathrm{min}}$, and the surface area in the filament. The dotted grey line indicates the introduction of $\rho_{\mathrm{min}}$ and corresponds to the top right panel of \cref{Fig: density_extended}. The dotted red line indicates the moment in which 1\%\ of the number of cells or the surface area has a density equal to $\rho_{\mathrm{min}}$ and corresponds to the bottom left panel of \cref{Fig: density_extended}. In the top panel of \cref{Fig: sa_sm_extended_run} we show that after the condensation is formed, its growth and increase in mass is quite steady. At most 77\%\ of the total mass is condensed into a cold dense filamentary structure. The fragmentation and introduction of a lower density bound does not substantially change the initial growth rate. After 1\%of the cells are modified, the growth stagnates. However, this almost coincides with the peak of fragmentation. The second panel of \cref{Fig: sa_sm_extended_run} shows that the additional mass induced by this measure is negligible until the 1\%\ mark. Afterwards, the additional mass increases to 5\%, at which point the simulation was terminated. The total number of cells or area with $\rho_{\mathrm{min}}$ also increases rapidly after the 1\%\ mark, reaching a maximum of 34\%. The behaviour and effect of the measure after the 1\%\ mark can be better understood by considering the bottom panels of \cref{Fig: density_extended}. In the moment in which the filament starts to break up into large strands, the growth in mass is stopped. From the pockets of the strands emerge long, white, low-density, regions. From these regions matter has been transported to the filament until $\rho_{\mathrm{min}}$ was reached. After breaking up, these large regions inject much additional mass, causing a continued increase in the total mass in the grid. Through the redistribution and interaction of the strands, the number of cells with $\rho_{\mathrm{min}}$ stops increasing. The surface area of the filament and strands is also almost constant during the redistribution and has a maximum of only 24\%\ of the grid. Hence, while the bootstrapping indeed allows us to continue far into the non-linear regime of the fragmentation of our filament, the effect of its continued mass addition is certainly noticeable.

\section{Conclusions and discussion}\label{Sec: sum-dis}

Optically thin radiative cooling is one of the important basic processes in physics. The precomputed, tabulated prescriptions or cooling curves vary widely due to differences in computations, atomic physics parameters, and assumed solar abundances \citepads{1978ApJ...220..643R}. In the literature they are mostly used as unquestioned truthful ingredients, but their effect on the simulations is not considered.  

We investigated how the condensation formation process is affected by the choice of cooling curve, and hence by the physical details of the curve. We used thermal instability as the mechanism in a simplified setup. For all tested cooling curves, the non-linear phase of the thermal instability led to the formation of a condensation that became a filament due to ram pressure. However, the growth rate of thermal instability is different for the various cooling curves. The growth rate obviously depends on the cooling rate at the initial temperature of 10$^6$ K, which is dependent on the abundances. Cooling curves based on older values for the abundances tend to underestimate the cooling rate because of the lower values for the abundances of low first-ionisation potential elements \citepads{1992PhyS...46..202F}. The second difference is in the morphology of the filaments that form. In some cases, the filament is much smaller in the direction perpendicular to the magnetic field. This is because the onset of fragmentation, which varies with respect to its growth rate for the thermal instability, is a determining factor. The shapes during fragmentation are also different. This might be because the seeds for the thin-shell instabilities, that is, small variations in the physical parameters, are slightly different as a result of the earlier linear evolution. We also found that some cooling curves are more stable against these instabilities, and that this is affected by the low (lower than 20\,000 K) temperature treatment. This is especially the case when the cooling curve is artificially caused to vanish at low temperatures. The fact that the choice of cooling curve and low-temperature treatment affects the dynamical stability might also be important in more complicated simulations of prominences and coronal rain, where it might be easier to form a stable prominence, but harder to form transient coronal rain using a certain cooling curve.

Furthermore, we have shown the need for high-resolution simulations by investigating the effect of varying the resolution. In our highest-resolution simulations, the dynamic behaviour of the condensation during formation and fragmentation occurred on length scales much finer than currently observable. Upcoming instruments such as DKIST \citepads{2020SoPh..295..172R}, which are able to resolve down to several dozen kilometres, are a great step forward.

Unfortunately, we noted that even at our highest resolution, we still encountered extremely low-density values in between strands of the fragmenting filament, similar to what has been observed in simulations by \citetads{2020A&A...636A.112C}. These cells become void because of the accretion onto the filament. The region heats up, but keeps the pressure steady. It can therefore not regain mass by dynamical means, that is, pressure gradients. Thermal conduction was not included in the simulations and might help solve this problem. In the cases of \citetads{2020A&A...636A.112C}, thermal conduction is present. However, the typical length scale for thermal conduction, the Field length, was not resolved. The low-density values lead to unfeasible short time steps due to the CFL condition. We provided all details of a possible bootstrap procedure that allowed us to simulate far into the nonlinear regime. It involves the sudden introduction of a lowest allowed density value. The filament breaks up and is redistributed as long strands throughout the domain. This phase is likely the most relevant for the ultimate shapes seen in (filamentary structured) prominences. It would be exciting to find observational evidence for the change in the filament orientation we predict from our simulations. The same effects have been discussed in 3D settings in \citetads{2020A&A...636A.112C}, but continuing into the far non-linear regime will require a bootstrap approach as advocated here. The additional effects of anisotropic thermal conduction, and the use of a numerically cured TRAC (transition region thermal conduction), will also play a role in future 3D simulations.

\begin{acknowledgements}

We wish to thank the anonymous referee, for the constructive comments that improved the paper, and the editorial office. JH would like to thank Niels Claes for providing the initial setup and Jack Jenkins for the useful discussions. The visualisations were achieved using the open source software ParaView (paraview.
org), Python (python.org) and yt (yt-project.org). The resources and services used in this work were provided by the VSC (Flemish Supercomputer Center), funded by the Research Foundation - Flanders (FWO) and the Flemish Government. Both authors are supported by the ERC Advanced Grant PROMINENT and a joint FWO-NSFC grant G0E9619N. This project has received funding from the European Research Council (ERC) under the European Union’s Horizon 2020 research and innovation programme (grant agreement No. 833251 PROMINENT ERC-ADG 2018). This research is further supported by Internal funds KU Leuven, project C14/19/089 TRACESpace.

\end{acknowledgements}

\bibliographystyle{aa}
\bibliography{main.bib}

\appendix
\section{MPI-AMRVAC cooling curves}\label{App: ccs}

We present the optically thin radiative cooling curves implemented in MPI-AMRVAC. \cref{Tab: ccs} lists basic properties of the cooling curves and their references. There are four piece-wise power laws: `Hildner', `FM', `Klimchuk', and `Rosner'. All other cooling curves are interpolatable tables. The tables are extended to include a cooling prescription at temperatures beyond the upper temperature limits. The piece-wise power laws are extended by using their last segment. For the interpolatable tables, bremsstrahlung is used. All tables with the `\_DM' part are modified with the same low-temperature treatment as prescribed in \citetads{2009A&A...508..751S}, where an ionisation fraction of 10$^{-3}$ is used. The parameter $\Lambda_{max}$ is defined for temperatures lower than $10^7$ K because of the ever-increasing cooling rate due to bremsstrahlung at high temperatures. The radiative and non-linear radiative timescales $\tau_{r}$ and $\tau_{nl}$ are calculated using \cref{Eq: tau_r} and \cref{Eq: tau_nl}, respectively. A typical solar density of 2.34~10$^{-15}$ g~cm$^{-3}$ is used in these calculations. Not all tables always assume collisional ionisation equilibrium. The `ML' and `cloudy' cooling curves do not. We refer to the accompanying papers for more details about their computations.

\begin{table*}[h!]
\caption{Cooling curves implemented in MPI-AMRVAC with basic properties and references.}
\centering
\begin{tabularx}{\textwidth}{l|c|c|c|c|c|c|c|X}
\hline
\hline
Name & T$_{min}$ & T$_{max}$  & T at $\Lambda_{max}$ & $\Lambda_{max}$ & $\Lambda (10^6 \text{K})$ & $\tau_{r} (10^6 \text{K})$  & $\tau_{nl} (10^6 \text{K})$ &  Reference and notes \\
Unit & K & K & K & erg cm$^3$ s$^{-1}$ & erg cm$^3$ s$^{-1}$ & s &  & \\\hline 

cloudy\_ism  & $10^{1}$  &  $10^{9}$  &  $10^{5}$  &  $10^{-21.28}$  &  $10^{-21.87}$  &  2963  &  0.026  & {\citetads{1998PASP..110..761F}, ISM metallicity} \\
cloudy\_solar  & $10^{1}$  &  $10^{9}$  &  $10^{5}$  &  $10^{-21.25}$  &  $10^{-21.79}$  &  2431  &  0.029  & {\citetads{1998PASP..110..761F}, solar metallicity} \\
Colgan  & $10^{4.06}$  &  $10^{9.06}$  &  $10^{5.29}$  &  $10^{-21.09}$  &  $10^{-21.31}$  &  803  &  0.117  & \citetads{2008ApJ...689..585C} \\
Colgan\_DM  & $10^{1}$  &  $10^{9.06}$  &  $10^{5.29}$  &  $10^{-21.09}$  &  $10^{-21.31}$  &  803  &  0.117  & \citetads{2008ApJ...689..585C} \\
Dere\_corona  & $10^{4}$  &  $10^{9}$  &  $10^{5.35}$  &  $10^{-21.18}$  &  $10^{-21.4}$  &  1008  &  0.135  & {\citetads{2009A&A...498..915D}, coronal abundances} \\
Dere\_corona\_DM  & $10^{1}$  &  $10^{9}$  &  $10^{5.35}$  &  $10^{-21.18}$  &  $10^{-21.4}$  &  1008  &  0.135  & {\citetads{2009A&A...498..915D}, coronal abundances} \\
Dere\_photo  & $10^{4}$  &  $10^{9}$  &  $10^{5.35}$  &  $10^{-21.25}$  &  $10^{-21.93}$  &  3364  &  0.047  &  {\citetads{2009A&A...498..915D}, photospheric abundances} \\
Dere\_photo\_DM  & $10^{1}$  &  $10^{9}$  &  $10^{5.35}$  &  $10^{-21.25}$  &  $10^{-21.93}$  &  3364  &  0.047  &  {\citetads{2009A&A...498..915D}, photospheric abundances} \\
DM  & $10^{2}$  &  $10^{9}$  &  $10^{5.3}$  &  $10^{-21.06}$  &  $10^{-22.28}$  &  7411  &  0.012  & {\citetads{1972ARA&A..10..375D}} \\
FM  & $10^{3}$  &  $10^{10}$  &  $10^{5.6}$  &  $10^{-21.06}$  &  $10^{-21.26}$  &  720  &  0.253  & {\citetads{1974SoPh...35..123H}, modified by \citetads{1991SoPh..135..361F}}\\
Hildner  & $10^{3}$  &  $10^{8}$  &  $10^{4.9}$  &  $10^{-21.1}$  &  $10^{-22.28}$  &  7204  &  0.005  & {\citetads{1974SoPh...35..123H}}\\
JCcorona  & $10^{4}$  &  $10^{7.95}$  &  $10^{5.29}$  &  $10^{-21.09}$  &  $10^{-21.31}$  &  803  &  0.117  & {\citetads{2008ApJ...689..585C}, modified by \citetads{2011ApJ...737...27X}} \\
Klimchuk  & $10^{3}$  &  $10^{8}$  &  $10^{4.97}$  &  $10^{-21.02}$  &  $10^{-21.72}$  &  2089  &  0.019  & {\citetads{2008ApJ...682.1351K}}\\
MB  & $10^{4}$  &  $10^{9}$  &  $10^{5.3}$  &  $10^{-21.42}$  &  $10^{-22.0}$  &  3969  &  0.052  & {\citetads{1981MNRAS.197..995M}}\\
MLcosmol  & $10^{2}$  &  $10^{9}$  &  $10^{4.2}$  &  $10^{-22.19}$  &  $10^{-23.51}$  &  127268  &  0.001  & {\citetads{2002A&A...394..901M}, zero metallicity} \\
MLsolar1  & $10^{2}$  &  $10^{9}$  &  $10^{5.4}$  &  $10^{-21.17}$  &  $10^{-22.2}$  &  6192  &  0.024  & {\citetads{2002A&A...394..901M}, solar metallicity} \\
MLwc  & $10^{2}$  &  $10^{9}$  &  $10^{5}$  &  $10^{-18.46}$  &  $10^{-20.23}$  &  68  &  0.002  & {\citetads{2002A&A...394..901M}, WC-star metallicity} \\
Rosner  & $10^{3}$  &  $10^{8}$  &  $10^{5.4}$  &  $10^{-21.2}$  &  $10^{-21.94}$  &  3457 &  0.046  & {\citetads{1978ApJ...220..643R}, extended by \citetads{1982soma.book.....P}} \\
SPEX  & $10^{3.8}$  &  $10^{8.16}$  &  $10^{5.32}$  &  $10^{-20.67}$  &  $10^{-21.71}$  &  2023  &  0.019  &  {\citetads{2009A&A...508..751S}} \\
SPEX\_DM  & $10^{1}$  &  $10^{8.16}$  &  $10^{5.32}$  &  $10^{-20.67}$  &  $10^{-21.71}$  &  2023  &  0.019  & {\citetads{2009A&A...508..751S}} \\
\hline
\end{tabularx}
\label{Tab: ccs}
\end{table*}

\section{Hydrodynamics (HD) case}\label{App: HD}

\begin{figure}[htbp]
  \resizebox{\hsize}{!}{\includegraphics{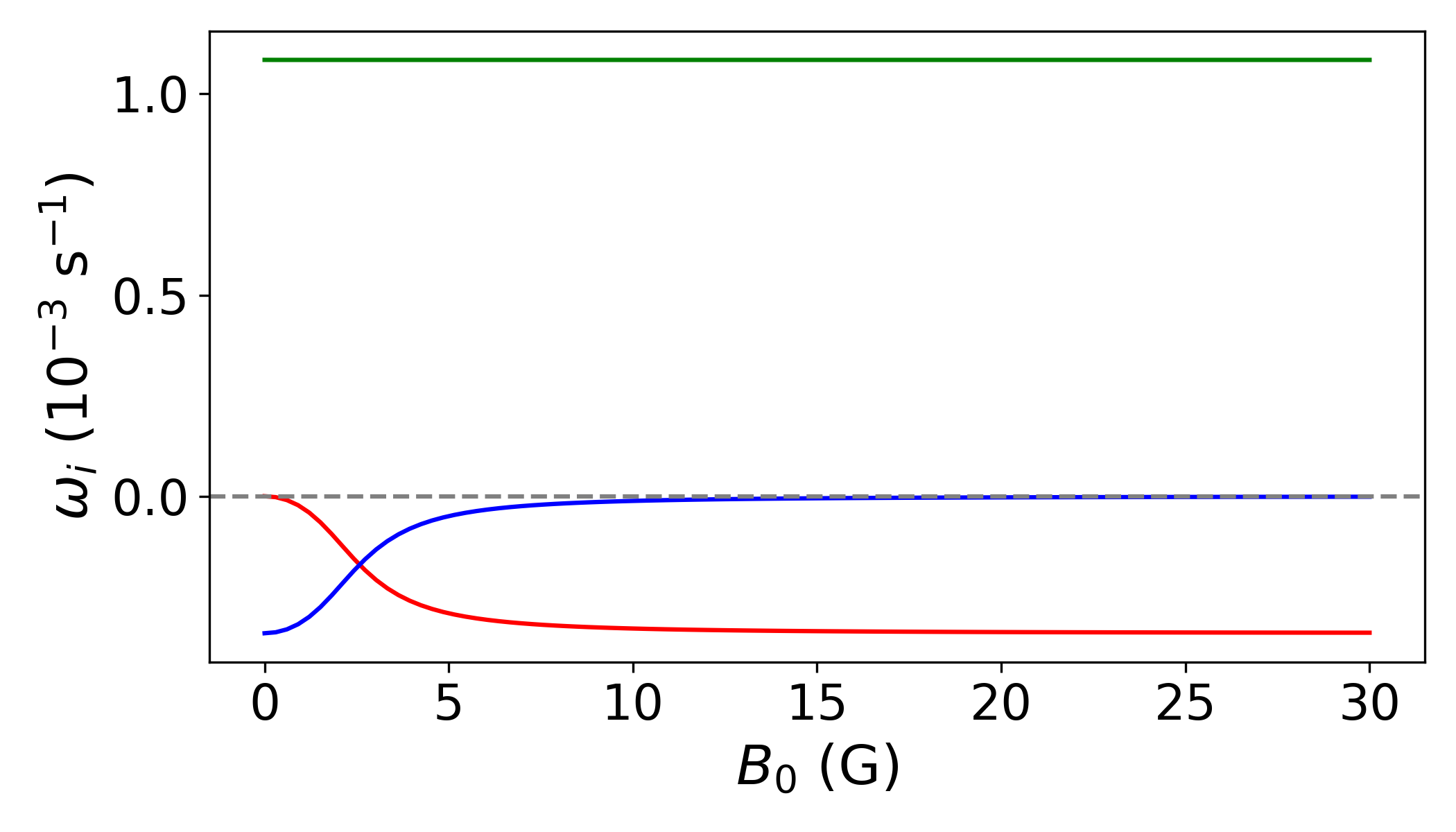}}
  \caption{Isochoric linear growth rates as a function of the background magnetic field $B_0$. The thermal, slow, and fast mode growth rates are denoted by the green, red, and blue lines, respectively.}
  \label{Fig: omegaIvsB}
\end{figure}

Some differences between thermal instability in hydrodynamics and magnetohydrodynamics are highlighted in this section. \cref{Fig: omegaIvsB} shows the isochoric growth rate, that is, the imaginary part of the complex eigenfrequency $\omega$ in our adopted $\exp{-i\omega t}$ temporal variation, of the thermal, slow, and fast MHD modes in their dependence on the background magnetic field strength. The plot was obtained by solving the isochoric dispersion relation for a uniform infinite medium, as can be found in \citetads{2019A&A...624A..96C}. The medium corresponds to the setup used in this manuscript with the background values given in \cref{Tab: initial_state}. The wave number and the angle between the background magnetic field and the wave vector were kept fixed to $2\pi$ divided by the domain length $L$ and $\pi/4$, respectively. 

We used the SPEX\_DM cooling curve and the background heating $H$, given by \cref{Eq: heating}. We recall that $\mathcal{L}$ is density dependent and given by \cref{Eq: curlyL}. The background magnetic field was varied from 0 to 30 G. At this angle and physical medium, the growth rate of the thermal mode remains constant for all background magnetic field strengths, as shown by the green line. This is not the case for the slow and fast MHD modes, which are shown by the red and blue lines, respectively. The slow MHD mode becomes less damped as the background magnetic field decreases. The mode completely vanishes in the hydrodynamical limit, $B_0$ = 0 G, as shown in the real part of the wave number in \cref{Eq: wr}. The damping rate of the fast MHD mode increases as the background magnetic field decreases. This mode has a sound wave character in the hydrodynamic limit.

To qualitatively show the difference between the result of a hydrodynamics and a low plasma beta magnetohydrodynamics thermal instability case, we modified the setup described in \cref{Sec: Setup}. The values of the background physical parameters were kept the same, but we omitted the magnetic field completely. We used the SPEX\_DM cooling curve. We also initiated the setup with two sound waves instead of slow MHD waves because there is no hydrodynamic analogue of the slow MHD wave.  \cref{Fig: omegaIvsB} shows that the role of slow and fast MHD waves changes drastically when the plasma beta changes to values below 1 (typical for our solar corona application). 

\begin{figure}[htbp]
  \resizebox{\hsize}{!}{\includegraphics{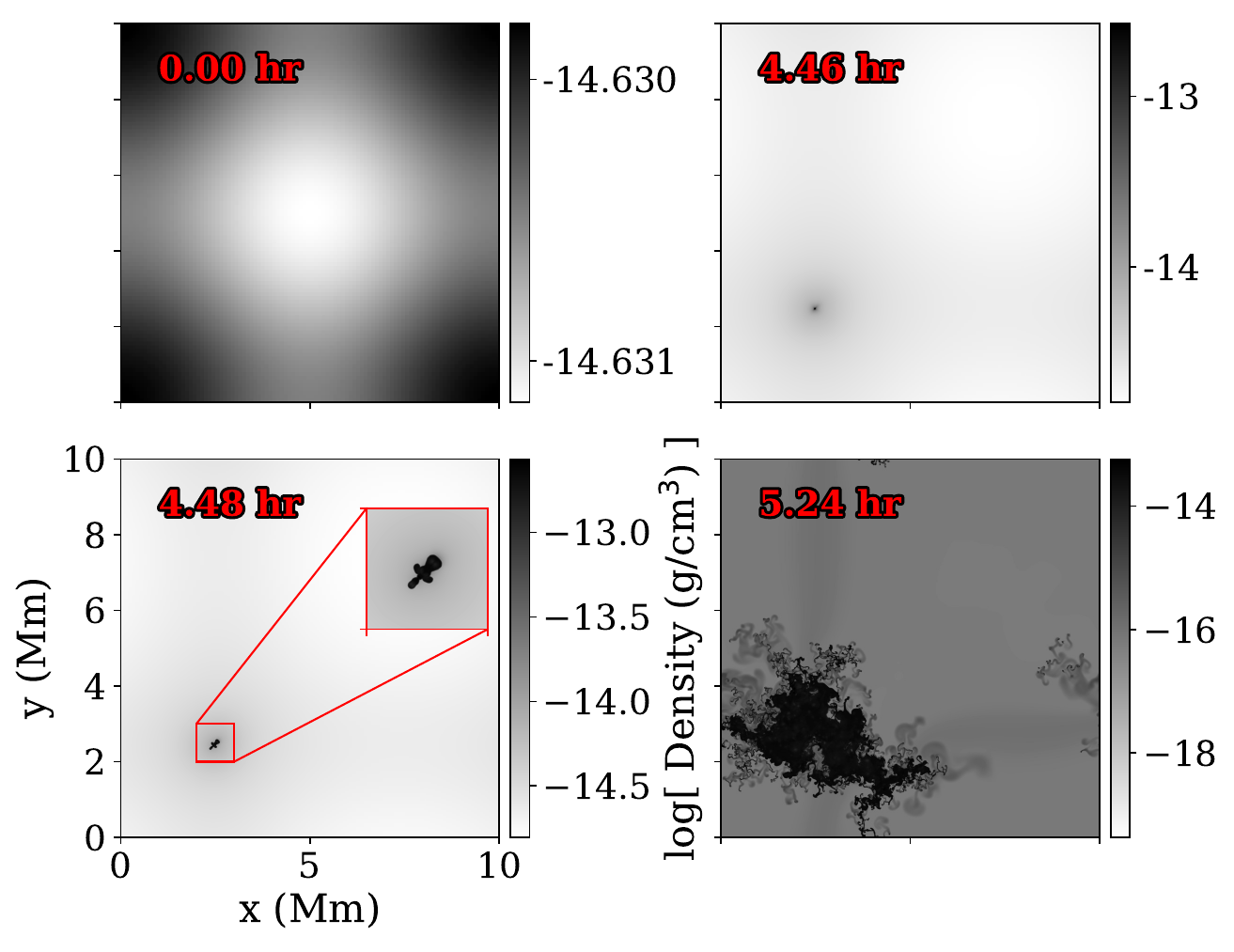}}
  \caption{Density view during the hydrodynamical case. \textit{Top left and right}: Initial state and onset of condensation formation, respectively. \textit{Bottom left and right}: Rapid expansion and fragmentation of the condensation, respectively. The animation of this simulation is available at \url{https://erc-prominent.github.io/team/jorishermans/}}
  \label{Fig: density_HD}
\end{figure}

\begin{figure}[htbp]
  \resizebox{\hsize}{!}{\includegraphics{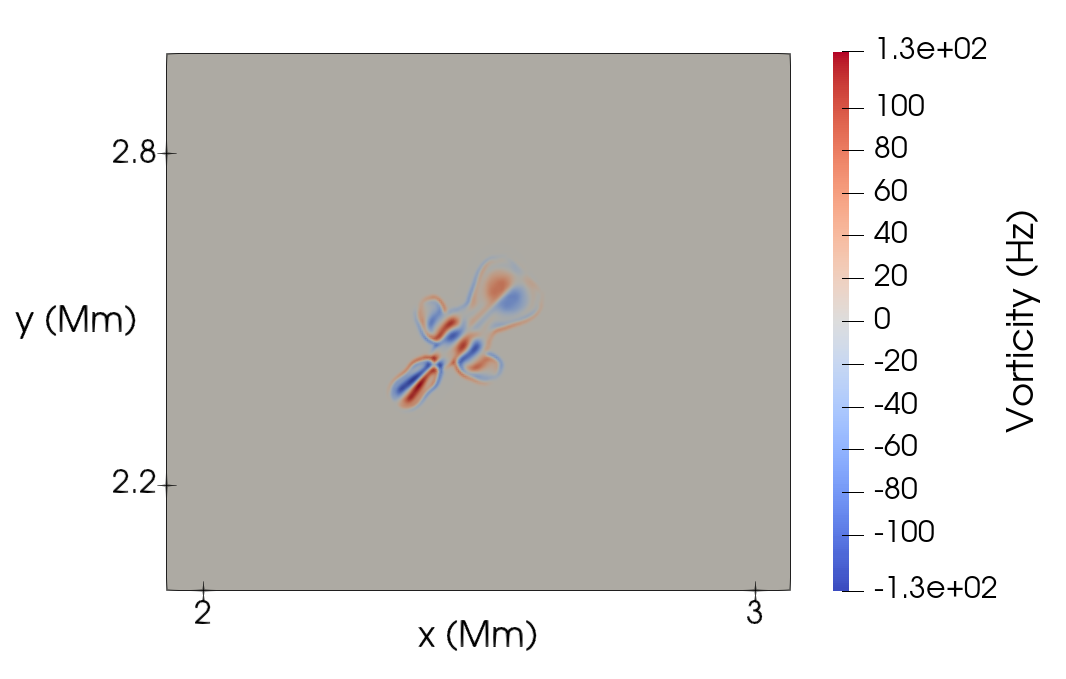}}
  \caption{Vorticity magnitude at the same moment as in the bottom left panel of \cref{Fig: density_HD}, zoomed onto the condensation.}
  \label{Fig: vorticity_HD}
\end{figure}

In \cref{Fig: density_HD} the density views are shown. The top left panel depicts the initial density state, which is not very different from the MHD case.
The top right panel shows the formation of the condensation after the non-linear growth of the density due to the thermal instability. The timescale at which this occurs is slightly different from the one of the MHD case because the sound waves are damped with a different damping rate than the slow MHD waves. The condensation that forms is nicely spherical symmetric because of the uniform inflow of matter, that is, there is no preferential direction because there is no magnetic field. There is only the fact that the combined sound wave pattern travels along the diagonal. Afterwards, the condensation fragments in all directions, as shown in the bottom panels. In the HD case the fragmentation is not the same as in the MHD case, and it is not caused by ram pressure differences. There is no consensus about the mechanism for fragmentation of high beta hydrodynamic gas clouds \citepads[see e.g.][]{2019ApJ...876L...3W,2019ApJ...875..158W,2020MNRAS.494L..27G,2021MNRAS.505.5238J,2021MNRAS.502.4935D}. One of those processes is `shattering' \citepads{2018MNRAS.473.5407M},  where a cloud fragments into small pieces that each cool isobarically, independently of each other. Because our result resembles the hydrodynamic simulation of \citetads{2021MNRAS.505.5238J}, the fragmentation process is most likely similar, although the initial conditions are very different. They indicate two processes for which the condensation is shredded apart through high vorticity, and they are thus explained differently from the `shattering' mechanism of \citetads{2018MNRAS.473.5407M}. In the first process, vorticity is generated in a rapid expansion phase by a Richtmyer-Meshkov \citepads{https://doi.org/10.1002/cpa.3160130207,1972FlDy....4..101M} process before the condensation fragments into smaller pieces \citepads{2020MNRAS.494L..27G}. In the second process, smaller condensations are fragmented by pressure gradients between merging large clouds \citepads{2019ApJ...876L...3W}. We also find an expansion phase before fragmentation in our simulation, as shown in \cref{Fig: density_HD}. The high vorticity is also apparent in \cref{Fig: vorticity_HD} at the same moment as in the bottom left panel of \cref{Fig: density_HD}. We did not include thermal conduction, and thus no evaporation, in our simulation.

Using equations 6 and 8 from \citetads{2019ApJ...875..158W}, we quantified the mode behaviour of our HD case, shown in \cref{Fig: density_HD}. Using their notation, we have $R=2.689$, $N_{\rho}=-0.575,$ and $N_P=-2.575.$ The setup is in case 6, where the plasma is both isobaric and isochoric unstable. In this regime, we indeed find two damped and propagating sound waves, and the unstable entropy mode responsible for thermal instability. The transition from strong background magnetic field and low plasma beta regime to the hydrodynamic limit and high plasma beta regime is worth a more detailed investigation.

\begin{figure}[htbp]
  \resizebox{\hsize}{!}{\includegraphics{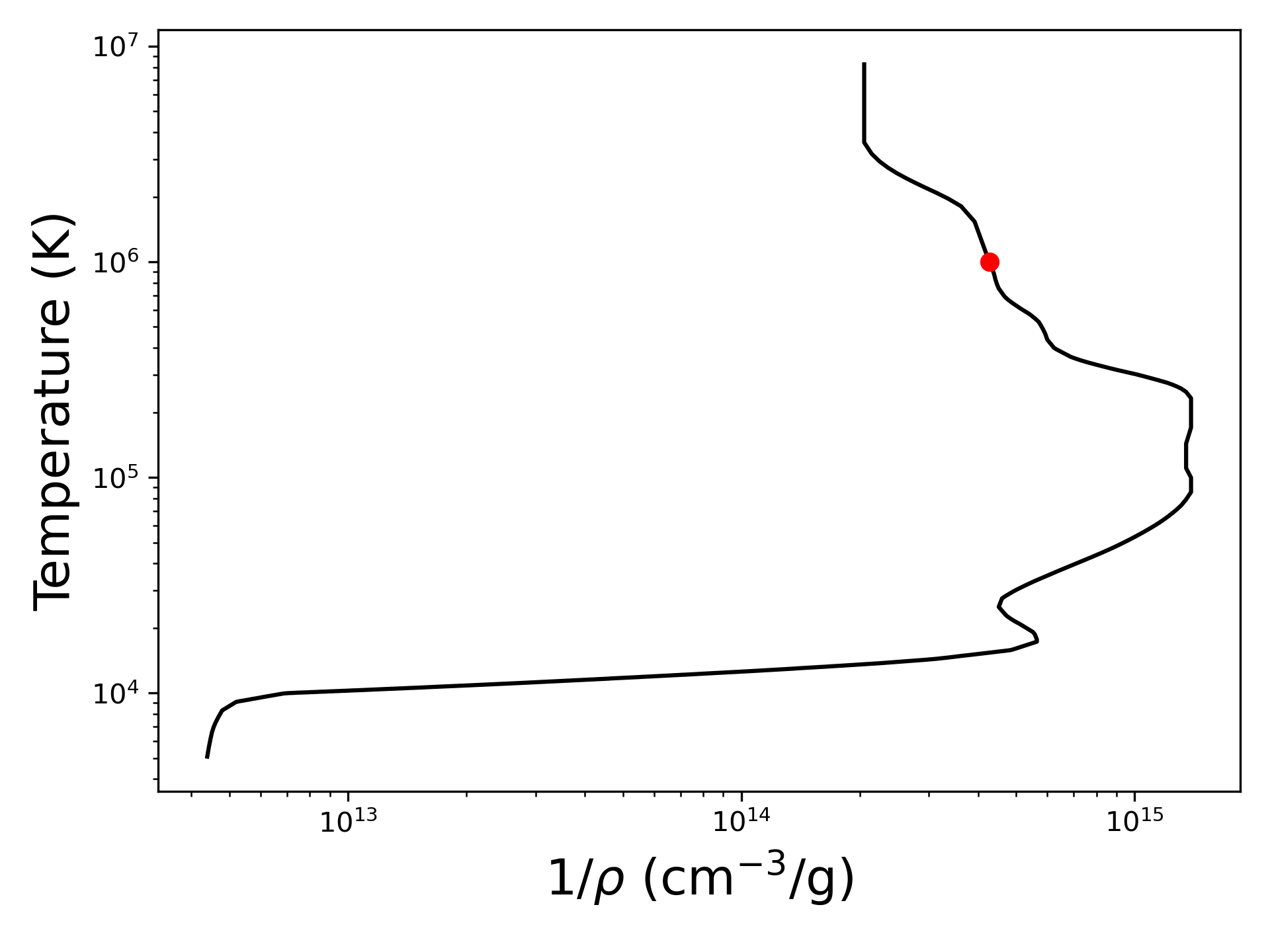}}
  \caption{Curve representing the solutions to the thermal equilibrium, analogously to the S-curves used by the AGN community. The black line represents all the solutions to the equilibrium $\mathcal{L}=0$ for the SPEX\_DM cooling curve and corresponding heating term. The red dot indicates the initial conditions.}
  \label{Fig: cc_WP}
\end{figure}

As an analogue to the S-curves used in the AGN and circumgalactic medium (CGM) community \citepads{1985ApJ...288...58L}, we can plot the information of the cooling and heating functions differently. We solved $\mathcal{L} = 0$ with $\mathcal{L}$ given by \cref{Eq: curlyL} for different densities. Plotting the temperature against the reciprocal of the density yields \cref{Fig: cc_WP}. This figure clearly shows that there is far more detailed variation than the S-curve used for example by \citetads{2019ApJ...875..158W}.

\end{document}